\documentstyle[12pt]{ioplppt}

\input{epsf}

\begin{document}

\jl{8}

\title{Entropy production, dynamical localization and criteria for quantum 
chaos in the open quantum kicked rotor}[Entropy, localization and quantum chaos]

\author{Paul A. Miller and Sarben Sarkar}

\address{Department of Physics, King's College London, Strand, 
London WC2R~2LS, UK}

\begin{abstract}

The von Neumann entropy production for a quantum mechanical kicked rotor 
coupled to a thermal environment is calculated. This rate of entropy increase
is shown to be a good criterion to distinguish between quantum mechanical 
counterparts of
chaotic and regular classical motion. We show that for high temperatures the 
entropy production rate increases linearly with the Kolmogorov-Sinai entropy 
of the classical system. However, for lower temperatures we also show that 
there are fluctuations in this linear behaviour due to dynamical localization.  

\end{abstract}


\maketitle

\section{Introduction}

The investigation of the quantum mechanical behaviour of systems with
classically chaotic counterparts is a rich and exciting field of
study. Many beautiful results have been found over the years regarding the 
semiclassical 
quantization of nonintegrable systems \cite{QC,OdeA,Gutz} and ``quantum 
chaology'' \cite{Haake,Houches,Berry1}. A major component of the approach 
involves the study of eigenvalue statistics, eigenfunction ``scarring'' etc. 
Classical chaotic behaviour, however, by definition a dynamical
phenomenon, seems to be suppressed at the quantum level \cite{Berry2,Ford}.
Apart from a brief time, during which its behaviour effectively mimics the
classically chaotic dynamics, the quantized version of a bounded and
conservative classical system is quasiperiodic and, therefore, predictable 
\cite{Ilg,Casati}.
                     
Chaotic systems thus test our understanding of the
relationship between classical and quantum mechanics. If classical behaviour 
is not a naturally emergent property of such systems in the absence of explicit
measurements then it is natural to 
consider the effect of environments on the quantum system \cite{Sarben}.
Quantum mechanical interference and superpositions, the cause of the 
aforementioned discrepancy, can be shown to be destroyed on extremely short 
timescales, in tractable model systems, by opening these systems to
environmental influences \cite{Zurek1,Book}. Some features of classical 
behaviour can then emerge without having to take the semi-classical limit.

Zurek and Paz \cite{Zurek2} have used the exactly solvable, inverted harmonic 
oscillator to
model heuristically unstable chaotic motion near a hyperbolic point,
although this system is certainly not chaotic itself. The Hamiltonian for 
this model is
\begin{equation}
H(x,p) = \frac{p^{2}}{2} - \frac{\lambda^{2} x^{2}}{2}.
\label{IHO}
\end{equation}
The effect of the the environment on this system might be expected to provide 
clues to
the behaviour that might be observed if we were to open a chaotic system
in a similar manner. The oscillator can be described by a master
equation in the position basis. 
Choosing the bath as a collection of harmonic oscillators
in thermal equilibrium at high temperature, with an Ohmic spectral density
proportional to a {\em small} dissipation parameter,
simplifies this equation enormously. Analysing the time evolution of an 
initial minimum uncertainty
state using this equation, Zurek and Paz have found that the von Neumann 
entropy, $S(t)$,  
quickly begins to increase linearly in time with a rate set by the 
quantity, $\lambda$ in equation (\ref{IHO}),  
\begin{equation}
\dot{S}(t) \equiv h_{Q} = \lambda.
\label{ZPC}
\end{equation}

Recently, however, we have generalized this result \cite{paper2} by 
considering the entropy production for all temperatures and for 
non-negligible dissipation parameters, $\gamma_{{\it diss}}$. We found that 
equation (\ref{ZPC}) is modified to
\begin{equation}
\dot{S}(t) \equiv h_{Q} = \sqrt{\lambda^{2} + \gamma_{{\it diss}}^{2}} - 
\gamma_{{\it diss}},
\end{equation}
but reduces to it when $\gamma_{{\it diss}} \ll \lambda$, as required.      

As $\lambda$ is loosely analagous
to the Lyapunov exponent of a chaotic system, Zurek and Paz conjecture 
that the entropy production rate might 
differentiate between chaotic and regular systems, and that this rate will
also be determined by the (positive) Lyapunov exponent of the classical
system. From the Alekseev-Brudno theorem (see p.4 in \cite{QC}, also
\cite{Beck})
\begin{equation}
\lim_{t\to\infty} \frac{I(t)}{t} = h_{KS} = \sum \Lambda^{+},
\label{ModCon}
\end{equation}
where $I(t)$ is the Shannon entropy, $h_{KS}$ is the Kolmogorov-Sinai entropy -
a measure of the average rate at which a classical system will lose
information \cite{Beck} - and the sum is over all the local positive Lyapunov 
exponents, $\{\Lambda^{+}\}$, in phase space. From this definition, 
of course, it is clear that $h_{KS}$ is actually an entropy {\it rate}.  
Consequently we will modify the Zurek and Paz conjecture to the statement that
$\dot{S}(t)$ increases {\em linearly} with $h_{KS}$. This has the attractive
feature of dealing both in the classical and quantum mechanical cases with the
change of entropy in time. Furthermore, the equality in equation (\ref{ZPC}) 
has been
replaced with the weaker statement of linear increase. This is the conjecture
that we will consider throughout the rest of our discussion.

The idea that quantum systems can be defined as ``chaotic'' or ``regular'', 
depending
on their behaviour under a perturbed evolution, is appealing and an old one 
\cite{Sarben}. 
An approach similar in spirit is due to Schack and Caves \cite{Caves}, who
have recently considered perturbed quantum maps in an information theoretic
framework and have established a criterion for chaos known as 
``hypersensitivity to perturbation''. 
In this paper we will study the prototypical, theoretically important
\cite{Reichl,LL} and 
experimentally realizable \cite{Moore,QKRdecoh} quantum kicked rotor. 
However, we will consider
the dynamics under the influence of a tractable, model environment and seek
to find evidence for the modified conjecture when the parameters of the 
combined system are 
compatible with those assumed during the framing of the conjecture.

The plan for the paper is as follows.
In Section \ref{section2} we shall briefly outline
the features of the classical and quantum kicked rotor relevant to the present
study.
In Section \ref{section3} we will derive the one-step propagators for both 
the (reduced)
density matrix of the rotor and its associated Wigner function representation.
The latter result allows us to see clearly that by opening the system we still
have a quantum analogue of the classical kicked rotor, or rather its standard
map. Also in this section we endeavour to quantify the effect of the 
environment on the coherence of the system. 

The results from our numerical analysis follow in Section \ref{section4} 
wherein we will
produce firm evidence in favour of the use of entropy production as a criterion
for quantum chaos {\em and} of the modified conjecture, at least in the 
parameter ranges
considered. We will also suggest explanations for those features found not
to be in accordance with the conjecture.
The conclusions follow in Section \ref{section5}.

\section{The Kicked Rotor}
\label{section2}

The kicked rotor model has been used extensively in studies of
classical chaos and its quantum manifestations 
\cite{Casati,Sarben,Reichl,LL,Chirik,Fox}.
Consequently the literature 
is vast, but for convenience we outline here those features of its 
complex behaviour
necessary for an understanding of our model and results. Even for readers 
familiar with the model a perusal would be useful in order to establish
the notation used.

\subsection{Classical Features}

The system Hamiltonian with which we will deal is that of a periodically kicked
rotor,
\begin{equation}
H_{rotor} = \frac{p^{2}}{2}-\frac{K}{4 \pi^{2}} \cos(2 \pi q) \sum_{n = 
- \infty}^{+ \infty} \delta (t-n),
\label{qkrham}
\end{equation}
where $p$ is an angular momentum and $q$ is a scaled angle variable.
Integrating the equations of motion from immediately after a delta function
"kick" to just after the next such kick will give a version of the famous
{\em standard map}
\begin{equation}
\eqalign{p_{n+1} = p_{n} - \frac{K}{2 \pi} \sin 2 \pi q_{n+1}, \\
q_{n+1} = (q_{n} + p_{n}) \pmod{1}.}
\label{classmap}
\end{equation}
The map is periodic, with period 1, in both directions of phase space - in
the angle direction by definition and in the momentum direction due to the form
of the map. This
means that one can simply view trajectories projected back onto the unit
square, $ \left[0,1 \right) \times \left[0,1 \right) $, in order to study
the dynamics for different initial conditions. However, even though there 
is periodicity in the momentum direction, the momentum is actually 
defined on the infinite interval $\left[ -\infty, \infty \right]$. The phase 
space has, therefore, the topology of a cylinder.  

The nonlinearity of the map is controlled by strength of the kick, $K$.
The trivial, regular motion when $K = 0$, becomes more and more
complicated as $K$ is increased \cite{Reichl,LL}. When $K < K_{c} \approx 
0.97\ldots$ some
KAM tori span the cylinder and prevent diffusion in the momentum direction. 
As $K$ increases still further these tori are destroyed, allowing the momentum
to diffuse, i.e.  
\begin{equation}
\langle (p(n) - p(0))^{2} \rangle = D(K) n,
\end{equation}
at a rate determined by the diffusion constant, $D(K)$. Here, the averaging 
is over a distribution of initial conditions. The diffusion constant can be 
calculated analytically \cite{LL,Rech}, and it is found that correlations in 
the angle variable from kick to kick influence its form considerably. If we 
ignore such correlations - an increasingly accurate approximation as 
$K \rightarrow \infty$ - the
angle variable can be assumed to be randomly but uniformly distributed on the 
unit interval. This {\it quasilinear} approximation allows us to write  
\begin{equation}
D_{{\it ql}}(K) = \frac{1}{2} \left(\frac{K}{2 \pi}\right)^{2}.
\label{Dql}
\end{equation}
Inclusion of these dynamical correlations leads to corrections which decrease 
in magnitude as $K$ gets larger, i.e. it is found \cite{LL,Rech} 
\begin{equation}
D(K) = \frac{1}{2} \left(\frac{K}{2 \pi}\right)^{2} \left( 1 - 
2 J_{2}\left(\frac{K}{2 \pi}\right) + 2 \left( J_{2}\left(\frac{K}{2 \pi}
\right) \right)^{2} + \ldots \right),
\label{DK}
\end{equation}
where $J_{2}$ denotes a Bessel function of order 2.   

For all finite values of $K$, KAM tori remain around elliptic points in 
phase space, of which there are an infinite number.  
These stable islands, surrounded by a chaotic sea, characterize what is 
known as a ``mixed" phase space, meaning that two nearby trajectories either
remain ``near" to one another for all time or diverge exponentially,
depending on whether they are to be found in a regular region or a chaotic one,
respectively (See figures [\ref{fig1}] and [\ref{fig2}]). Hence the Lyapunov 
exponents for the system are dependent on the starting point in phase space, 
but gradually converge to a uniform value as $K \rightarrow \infty$. 

From Pesin's theorem \cite{Pesin}, the kicked rotor, being a chaotic system 
with one degree-of-freedom,
possesses a positive {\em KS entropy}, $h_{KS}$, which will be equal to its 
positive Lyapunov exponent for $K$ sufficiently large. Chirikov \cite{Chirik}
has calculated this quantity and found
\begin{equation}
h_{KS} \approx \ln \left(\frac{K}{2} \right).
\label{KSent}
\end{equation}

\subsection{Quantum Features}

The periodicity of the angle variable gives rise to a numerable and discrete
set of momentum eigenvalues defined by
\begin{equation}
\hat{p} |l \rangle = p_{l} |l \rangle = 2 \pi \hbar l |l \rangle,
\end{equation}
where $l$ is an integer. The corresponding eigenstates are obviously
eigenstates too of the kinetic energy operator $\hat{H}_{0} = \hat{p}^{2} / 2$, 
and we write
\begin{equation}
\hat{H}_{0} |l \rangle = E_{l} |l \rangle = 2 \pi^{2} \hbar^{2} l^{2} |l 
\rangle.
\end{equation}

The Hamiltonian of equation (\ref{qkrham}) leads to a one-step unitary
transformation
\begin{equation}
U(t_{n+1},t_{n}) = U_{k} U_{f},
\label{Ut}
\end{equation}
which defines the quantum standard map. Here, $t_{n}$ is the time after the
$n$th kick and $U_{k}$ is the unitary kick evolution operator. It can be 
written in the momentum representation as
\begin{equation}
\langle l^{\prime}| U_{k} | l \rangle = i^{l^{\prime}-l} J_{l^{\prime}-l}
(K/4 \pi^{2}\hbar),
\label{Uk}
\end{equation}
with $J_{m}$ denoting a Bessel function of integer order $m$. $U_{f}$ 
is the operator corresponding to unitary free rotation between
the delta function kicks, and has the momentum representation
\begin{equation} 
\langle l^{\prime}| U_{f} | l \rangle = \exp ( -i 2 \pi^{2} \hbar l^{2})
\, \delta_{l^{\prime}l}.
\label{Uf}
\end{equation}

The unitary dynamics was first investigated by Casati at al. \cite{Casati} 
using initial states highly localized in momentum. It was found that 
$\langle (\hat{p}(n) - \hat{p}(0)) ^{2} \rangle$ shows diffusive 
behaviour in approximate agreement with the classical case, but only up to a 
finite {\it break} time, $n^{*}$, after which the momentum
ceases to grow further and quasiperiodic motion around some finite average is
found. This defines a {\it localization length} via 
\begin{equation}
\overline{\langle (\hat{p}(n) - \hat{p}(0)) ^{2} \rangle} = (2 \pi \hbar)^{2} 
L^{2},
\end{equation}
for $n > n^{*}$. Good analytic and numerical evidence now exists pointing to 
the fact that the eigenstates of the unitary evolution operator of 
equation (\ref{Ut}) above are localized in momentum and have essentially 
discrete 
spectra on the unit circle of the complex plane. These assumptions lead to 
the following estimates for $L$ and $n^{*}$ \cite{DG1} 
\begin{equation} 
L \approx \left( \frac{D(K)}{2 \pi^{2} \hbar^{2}} \right),
\label{loclength}
\end{equation}
and
\begin{equation}
n^{*} \approx 2L.
\label{breaktime}
\end{equation}
It is important to note that both quantities are inversely proportional to 
the square of the 
scaled unit of action, $\hbar$. We also note that {\it classical} correlations, 
which can both increase and decrease the diffusion constant from its
quasilinear value, $D_{{\it ql}}$, (see equations (\ref{Dql}) and (\ref{DK})) 
can also directly affect the quantum dynamics via this localization mechanism.  

The discrepancy between the classical and quantum diffusive 
behaviour of the kicked rotor is therefore known to be a consequence of 
quantum interference. If the
phase coherence could somehow be destroyed one might expect to recover the 
classical diffusion. Various authors have considered this problem 
\cite{Sarben,Ott,Cohen,DG1} and have
examined the effect of an environment on localization. However, previously the
 relationship between the classical and quantum entropy 
production, the former being a quantitative measure of chaos, has {\em not} 
been studied. This approach is important since we have growing evidence that 
it is a generic one.

\section{Open Dynamics}
\label{section3}

We now proceed to derive the propagator for the density matrix of 
the rotor after tracing out the environmental degrees of freedom, i.e. the
propagator for the {\em reduced} density matrix. This will enable us to 
calculate
the von Neumann entropy at each time step, upon iteration of an arbitrary
initial state.

\subsection{The Reduced Density Matrix Propagator}

We will restrict attention to the situation {\em between} kicks initially and
consider the free rotor to be coupled through its momentum to a thermal bath 
of harmonic oscillators. 
The Hamiltonian for this system,
\begin{equation}
H = \frac{p^{2}}{2} + \sum_{k} ( \frac{p_{k}^{2}}{2 m_{k}} + \frac{m_{k}
\omega_{k}^{2} q_{k}^{2}}{2} ) + p \sum_{k} c_{k} q_{k},
\end{equation}
can be written as
\begin{equation}
H = \frac{p^{2}}{2} + \sum_{k} \hbar \omega_{k} (b_{k}^{\dagger} b_{k} +
\frac{1}{2})
+ p \sum_{k} \overline{c_{k}}(b_{k}^{\dagger} + b_{k}),
\end{equation}
where $\overline{c_{k}} = \sqrt{\hbar / (2 m_{k} \omega_{k})}c_{k}$
is a renormalized coupling and
\begin{eqnarray}
b_{k} & = & \sqrt{\frac{m_{k} \omega_{k}}{2 \hbar}} \left( q_{k} + i\frac{p_{k}}
{m_{k} \omega_{k}} \right), \nonumber \\
b_{k}^{\dagger} & = & \sqrt{\frac{m_{k} \omega_{k}}{2 \hbar}} \left( q_{k} - 
i\frac{p_{k}}{m_{k} \omega_{k}} \right). \nonumber
\end{eqnarray}
As $[ q_{j}, p_{k} ] = i \hbar \delta_{jk}$, these annihilation and creation 
operators satisfy $[ b_{j}, b_{k}^{\dagger} ] = \delta_{jk}$.   

We choose an uncorrelated initial product state
\begin{equation}
\rho(0) = \rho^{S}(0) \otimes \prod_{k} \rho_{th}^{k},
\end{equation}
where $\rho_{th}^{k}$ is the density matrix of the $k$th harmonic oscillator
in thermal equilibrium at temperature $T$, defined by
\begin{equation}
\rho_{th}^{k} = \left[ 1 - \exp \left(- \frac{\hbar \omega_{k}}{k_{B} T}\right) 
\right]^{-1} \sum_{n}
\exp \left( \frac{- n \hbar \omega_{k}}{k_{B} T} \right) |n \rangle \langle n|.
\end{equation}

To specify the problem further we also need to choose a physically reasonable
environmental {\em spectral density} \cite{Leggett}, $I(\omega)$, defined 
formally by
\begin{equation}
I(\omega) = \sum_{k} \delta(\omega - \omega_{k}) \frac{c_{k}^{2}}{2 m_{k}
\omega_{k}}.
\end{equation}
More generally, if we assume every environmental oscillator to have equal mass,
 we can write \cite{Book}
\begin{equation}
I(\omega) \approx S(\omega) \frac{c^{2}(\omega)}{2 m \omega},
\end{equation}
with a smooth density $S$. In order to avoid the influence of 
unphysical, extremely high frequencies we assume the existence of an upper,
or cutoff frequency, $\omega_{c}$, and choose a continuum of oscillators
distributed according to the density
\begin{equation}
S(\omega)c^{2}(\omega) = 2 m \eta \omega^{q} \exp (-\frac{\omega}{\omega_{c}}),
\label{density}
\end{equation}
with $q > 1$, which reduces to that used by Caldeira and Leggett 
\cite{Leggett} when we choose a so-called {\it Ohmic} environment with 
$q = 2$.   
Hamiltonians and initial conditions of this Ohmic type have been considered by 
Jiushu et al. in 
\cite{jiushu} and the {\em exact} time evolution for the reduced density matrix
has been obtained. We proceed along similar lines but consider the general 
case with $q > 1$ being the only assumption. We arrive finally
at an expression for the reduced density matrix of the rotor in the momentum
representation
\begin{equation}
\fl \langle m |\rho^{S}(t)| n \rangle = \exp \left( -\frac{i}{\hbar} 
(E_{m}-E_{n})(t + A(q,t)) - \frac{(p_{m} - p_{n})^{2}}{\hbar} B(q,t) \right) 
\langle m |\rho^{S}(0)| n \rangle.  
\label{rdm}
\end{equation}
The environmental terms in this expression, given by $A(q,t)$ and 
$B(q,t)$, have been evaluated explicitly for all $q > 1$ \cite{thesis}. 
The expressions are particularly simple, however, for the Ohmic case, and 
we concentrate on this important case from here on. We find 
\begin{equation}
A(2,t) = 2 \eta (\arctan (\omega_{c} t) - \omega_{c} t),
\label{A}
\end{equation}
and
\begin{equation}
B(2,t) = \frac{ \eta}{2} \ln (1 + \omega_{c}^{2} t^{2}) + \eta \ln \prod_{j =
1}^{\infty}(1 + ( \frac{\omega_{c} t}{1 + j \beta \hbar \omega_{c}})^{2}),
\label{B}
\end{equation}
where the usual notation of $ \beta = 1/(k_{B} T) $ has also been used.

There are a few points to note here. Firstly, were $K = 0$ in
equation (\ref{qkrham}), then equation (\ref{rdm}) would describe fully the 
time evolution of any initial state,
 $ \rho^{S}(0) $, in the momentum representation, for a classically 
{\em nonchaotic} rotor. Secondly, we return to the correct unitary dynamics
, corresponding to conservative evolution of the rotor, when $\eta$ is set to
 $0$. (This amounts to setting all coupling constants, $c_{k}$, to zero.)
Finally, we will consider equation (\ref{rdm}) to describe the time evolution 
of the
 quantum kicked rotor between kicks and so will set $t = 1$ in 
equations (\ref{rdm}), (\ref{A}) and (\ref{B}). The heat bath is assumed to be
sufficiently large to remain in thermal equilibrium as it interacts with the 
system between each kick. For this assumption to be valid 
the coupling constants, $ \{ c_{k} \} $, are taken to be {\em small}. 
Hence, for the spectral density in equation (\ref{density}), $\eta \ll 1$, a 
common assumption \cite{Zurek2,DG1}.
 
The system evolution, from just after the kick at $t = n$  
until just {\em before} the kick at $t = n+1$, where $n$ is an integer, 
is given by a master equation of the form   
\begin{equation}
\langle l^{\prime}| \rho^{-}(n+1) | m^{\prime}\rangle = \sum_{l,m}
G^{f}(l^{\prime},m^{\prime},l,m) \langle l | \rho^{+}(n) | m \rangle,
\label{fprop}
\end{equation}
where
\begin{equation}
G^{f}(l^{\prime},m^{\prime},l,m) = G^{f}_{C}(l^{\prime},m^{\prime},l,m)
\Gamma(l,m),
\label{Gf}
\end{equation}
$G^{f}_{C}(l^{\prime},m^{\prime},l,m)$ is the unitary summation kernel
given by
\begin{equation}
G^{f}_{C}(l^{\prime},m^{\prime},l,m) = \exp(-\frac{i}{\hbar} (E_{l}-E_{m}))
\, \delta_{l^{\prime}l} \delta_{m^{\prime}m}
\end{equation}
and
\begin{equation}                                                               
\Gamma(l,m) = \exp(- \frac{i}{\hbar} (E_{l}-E_{m}) A(2,1) - 
\frac{(p_{l} - p_{m})^{2}}{\hbar} B(2,1) )
\end{equation} 
describes the non-unitary influence of the environment on the dynamics.

It is assumed that during each kick both free evolution and the
influence of the environment are negligible. We are then led formally to the 
summation kernel for the kick, defined by the master equation
\begin{equation}
\langle l^{\prime}| \rho^{+}(n+1) | m^{\prime} \rangle = \sum_{l,m}
G^{k}(l^{\prime},m^{\prime},l,m) \langle l | \rho^{-}(n+1) | m \rangle.
\label{kprop}
\end{equation}
Choosing now the shorthand notation used by Dittrich and Graham \cite{DG3} of
$k := K/4 \pi^{2} \hbar$ and $b_{m}(x) := i^{m} J_{m}(x)$ we can write
explicitly 
\begin{equation}
G^{k}(l^{\prime},m^{\prime},l,m) = b_{l^{\prime}-l}(k) \,b^{*}_{m^{\prime}-m}
(k).
\label{Gk}
\end{equation}

Even as the density matrix evolves, according to master equations 
(\ref{fprop}) and (\ref{kprop}) above, it must, of course, remain normalized, 
hermitian and positive \cite{DG2}. These properties in turn impose conditions 
on the kernels of equations (\ref{Gf}) and (\ref{Gk}) which can 
easily be shown to hold \cite{thesis}, implying that the evolution of the 
reduced quantum statistical density operator is a valid one.

The one time-step evolution of the density matrix is now simply the convolution
of $G^{f}$ and $G^{k}$ defined in equations (\ref{Gf}) and (\ref{Gk}) above,
respectively, and it is easily confirmed that   
\begin{equation}
\langle l^{\prime}| \rho^{+}(n+1) | m^{\prime}\rangle = \sum_{l,m}
G(l^{\prime},m^{\prime},l,m) \langle l | \rho^{+}(n) | m \rangle,
\label{master}
\end{equation}
where
\begin{equation}
G(l^{\prime},m^{\prime},l,m) = G_{C}(l^{\prime},m^{\prime},l,m)
\Gamma(l,m),
\label{rhoprop}
\end{equation}
in which
\begin{equation}
G_{C}(l^{\prime},m^{\prime},l,m) = \langle l^{\prime} | U | l \rangle \, 
\langle m | U^{\dagger} | m^{\prime} \rangle
\end{equation} 
is the unperturbed, unitary propagator. When there is no environment, i.e. 
as $\eta \to 0$, then $\Gamma(l,m) \to 1$ for each $l$ and $m$.

\subsection{Destruction of Coherence}

It was stated above that the term $\Gamma(l,m)$ in equation (\ref{rhoprop}) 
makes a non-unitary contribution to the dynamics. In this section we 
would like to quantify this statement. 

A measure of decoherence is the decay probability, $P_{1}$, of a quasienergy 
eigenstate in a single timestep. In order to calculate $P_{1}$ we first 
consider the probability of transition, $P^{1}(s|r)$, from an quasienergy 
eigenstate $|r\rangle$ to a 
quasienergy eigenstate $|s\rangle$ in a single timestep. In fact, if 
$\rho^{+}(0) = |r\rangle \langle r|$ then we may write
\begin{equation}
P^{1}(s|r) = \langle s| \rho^{+}(1) |s \rangle.
\label{probsr1}
\end{equation}
The propagator in the quasienergy representation is therefore required, and 
the master equation (\ref{master}) maybe be rewritten
\begin{equation}
\langle \alpha_{1}| \rho^{+}(n+1) | \beta_{1}\rangle = 
\sum_{\alpha_{0},\beta_{0}}
G^{qe}(\alpha_{1},\beta_{1},\alpha_{0},\beta_{0}) 
\langle \alpha_{0} | \rho^{+}(n) | \beta_{0} \rangle,
\label{quasiprop}
\end{equation}
where 
\begin{equation}
U|\alpha_{0} \rangle = e^{i \phi_{\alpha_{0}}} |\alpha_{0} \rangle,
\end{equation}
and similarly for $|\alpha_{1} \rangle, | \beta_{0} \rangle$ and 
$| \beta_{1}\rangle$.
One easily finds 
\begin{equation}
G^{qe}(\alpha_{1},\beta_{1},\alpha_{0},\beta_{0}) = \sum_{l,m} 
e^{i \left( \phi_{\alpha_{1}} - \phi_{\beta_{1}} \right) }
\langle \alpha_{1} | l \rangle  \langle l | \alpha_{0} \rangle 
\langle \beta_{0} | m \rangle \langle m | \beta_{1} \rangle 
\Gamma(l,m).                                        
\end{equation}
Thus, because of equations (\ref{probsr1}) and (\ref{quasiprop}), we find 
\begin{eqnarray}
P^{1}(s|r) &=& G^{qe}(s,s,r,r) \nonumber \\
&=& \sum_{l,m}                                            
\langle s | l \rangle  \langle l | r \rangle 
\langle r | m \rangle \langle m | s \rangle \Gamma(l,m).
\label{probsr2}
\end{eqnarray}
We can now use this expression to calculate $P_{1}$ by summing over all
possible final states, $| s \rangle$ (different from $| r \rangle$) and 
averaging over the various possible initial states (see, for example, 
reference \cite{Cohen}): 
\begin{equation}
P_{1} = \frac{1}{D} \sum_{r,s \atop (r \not= s)} P^{1}(s|r),
\label{P1}
\end{equation}
where $D$ is the dimension of the truncated basis (See Section \ref{section4}). 
Inserting equation (\ref{probsr2}) and using the fact that
\begin{equation}
\sum_{s \atop (r \not= s)} \langle s | l \rangle \langle m | s \rangle 
= \delta_{m,l} - \langle r | l \rangle \langle m | r \rangle
\end{equation}
we find
\begin{equation}
P_{1} = 1 - \frac{1}{D} \sum_{r} \gamma_{r},
\end{equation}
where we have defined
\begin{equation}
\gamma_{r} = \sum_{l,m} |\langle r | l \rangle|^{2} 
|\langle r | m \rangle|^{2} \Gamma(l,m).
\end{equation}

We may estimate this expression analytically \cite{thesis} if we assume 
that {\it every} quasienergy eigenfunction is exponentially localized 
according to 
\begin{equation}
|\langle r | l \rangle|^{2} \approx \exp \left( - \frac{|l - l_{r}|}{L} 
\right).
\end{equation}
Here we have denoted the centre of localization by $l_{r}$ and $L$ is 
the localization length defined previously in equation (\ref{loclength}). 
However, not every such eigenstate has this structure, as certain 
``double-hump" quasienergy eigenstates exist, so-called because they have two 
localization centres. 

For accuracy therefore, we have calculated $P_{1}$ numerically and examined 
the effect changing $K$, $\hbar$ and $\beta$, the environmental inverse 
temperature. We present our results in Section \ref{section4}.

\subsection{The Wigner Distribution Propagator}

We would like now to examine the classical limit of our system. To do so we
choose to work in the Wigner representation for a phase space with the
topology of a cylinder. The
notation in this section is exactly that used in refs.\cite{DG2,DG3}, in the 
latter of which this representation is succinctly descibed. 

The Wigner distribution on our cylindrical phase space at time $n$,
$W_{n}(q,p)$, consists of $\delta$-functions concentrated on the definite 
values
of momentum, $p = \pi \hbar l$, for $l$ an integer. One therefore writes
\begin{equation}
W_{n}(q,p) = \sum_{l} W_{n}^{(l)}(q) \delta (p - \pi \hbar l)
\end{equation}
for the Wigner distribution at this time. Its (one) time-step
generator, $G_{W}$,  is defined by
\begin{equation}
W_{n+1}^{(\overline{l})}(\overline{q}) = \sum_{l} \int_{0}^{1} dq \, 
G_{W}(\overline{l},
\overline{q}, l, q) W_{n}^{(l)}(q),
\end{equation}
and can be expressed in terms of the one-step density matrix propagator, 
most conveniently in the momentum representation, using
\begin{eqnarray}
\fl G_{W}(\overline{l}, \overline{q}, l, q) = \sum_{m_{1},
m_{2}} \sum_{l_{1}, l_{2}} \, \delta_{\overline{l},m_{1}
 + m_{2}} \delta_{l, l_{1} + l_{2}} \exp \{ 2 \pi i [ (m_{1} - m_{2}) 
\overline{q} - (l_{1} - l_{2}) q ] \} \nonumber \\
\times G ( m_{1}, m_{2}, l_{1}, l_{2} ).
\label{rhotoW}
\end{eqnarray}
It will be convenient to make the transformation from the density matrix 
propagator to
the Wigner function propagator for both the free and the kicked part 
separately. A convolution of the two will then yield the desired quantized
map in the Wigner function representation.    
 
Use of equation (\ref{Gf}) in equation (\ref{rhotoW}) gives
\begin{eqnarray}
G^{f}_{W}(\overline{l},\overline{q}, l, q) & = & \delta_{\overline{l}l} 
\mbox{Ga}(x,\Delta,1), \\
 & \equiv & \delta_{\overline{l}l} \sum_{k = - \infty}^{\infty} \exp \left( 
2 \pi i k x - 2 \pi^{2} k^{2} \Delta \right), \nonumber
\end{eqnarray}
i.e., proportional to a periodic Gaussian with period 1. Here, $x$ and $\Delta$
 are defined by  
\begin{equation}
x = \overline{q} - q - \pi \hbar \overline{l} (1 + A(1)),                   
\end{equation}
and
\begin{equation}
\Delta = 2 \hbar B(1).
\end{equation}
This periodic Gaussian reduces to a periodic $\delta$-function with the 
argument $\overline{q} - q - \pi \hbar \overline{l} \,$ for $\eta \rightarrow 0$,
 as expected. 
Similarly, use of equation (\ref{Gk}) in equation (\ref{rhotoW}) gives
\begin{equation}
\fl G^{k}_{W}(\overline{\overline{l}},\overline{\overline{q}}, \overline{l}, 
\overline{q}) = i^{\overline{l}-\overline{\overline{l}}} \exp \{2 \pi i  
\overline{l} (\overline{q}-\overline{\overline{q}})\}
J_{\overline{l}-\overline{\overline{l}}}(2 k \sin(2 \pi 
\overline{\overline{q}})) \sum_{m} \exp \{ -4 \pi i m (\overline{q}-
\overline{\overline{q}})\}.
\end{equation}
There remains only to convolute the two propagators to give
\begin{eqnarray}
\fl G_{W}(\overline{\overline{l}},\overline{\overline{q}}, l, q) & = 
\sum_{\overline{l}} \int_{0}^{1} d\overline{q} \,
G^{k}_{W}(\overline{\overline{l}},\overline{\overline{q}},\overline{l},
\overline{q})
G^{f}_{W}(\overline{l},\overline{q}, l, q) \nonumber \\
 & = J_{\overline{\overline{l}}-l}(2 k \sin(2 \pi 
\overline{\overline{q}})) \, 
\mbox{Ga}(\overline{\overline{q}} - q - \pi \hbar l (1 + A(1)),2 
\hbar B(1),1).         
\end{eqnarray}

When $\eta \ll 1$ and $\hbar \rightarrow 0$ this quantum map can be shown
\cite{Jensen} to reduce to the classical map of equation (\ref{classmap}), 
as required.

\section{Results}

\label{section4}

We now present the results of our numerical calculations. For each run we
began with an initial coherent state centred at some
point $(p,q)$ in phase space. We choose Planck's constant,
$h$ to be a rational multiple of 
\begin{equation}
\gamma := \frac{1}{\sqrt{5} - 1}.
\end{equation}
This avoids any 
resonances \cite{Reichl}, which might occur between the frequency of the 
driving force and those frequencies associated with the 
energy levels, $\{E_{n}\}$, of the unperturbed, regular rotor for rational $h$. 
(Such resonances give rise to a superdiffusive, quadratic increase of the 
rotor's energy with time in contrast to the linear increase found for the
classical rotor.)  
The infinite
dimensionality of the Hilbert space forces us to truncate the momentum basis 
in which we work. We have chosen, for the generation of the results presented 
below, to use a truncated basis set which is symmetric about the zero momentum
eigenstate. That is, we have used the set $\{|-n \rangle, |-n+1 \rangle , 
\ldots, |-1
\rangle, |0 \rangle, |+1 \rangle, \ldots, |n-1 \rangle, |n \rangle\}$, where 
$n$ was chosen large enough to keep the density matrix normalized to an 
accuracy of better than 1 part in $10^{7}$; i.e.
\begin{equation}
\mbox{Tr} \left[ \rho_{S}(t) \right] = 1,
\end{equation}
for the time of interest. Typically, $n$ ranged from $\approx 250$ up to 
$\approx 450$, depending on the parameter values used. 

A propagation through one time step was then effected by computing 
equation (\ref{fprop})
followed by equation (\ref{kprop}). At each stage the von Neumann entropy, 
$S(t)$
 was
calculated by using the approximation
\begin{equation}
S(t) := - \mbox{Tr} \left[\rho_{S}(t) \ln \rho_{S}(t) \right] \approx - 
\sum_{i = -n}^{n} \lambda_{i} \ln \lambda_{i},
\end{equation}
where $\{ \lambda_{i}| -n \leq i \leq n \}$ is the set of eigenvalues of
the density matrix $\rho_{S}(t)$ at that time.
The entropy is a maximum if and when each $\lambda_{i} = 1/(2n+1)$, giving
an upper limit to the entropy of $S_{max} = \ln (2n+1)$ as a 
result of the truncation.

Finally, as we wish to investigate the modified conjecture it is reasonable 
to make the same characterisation for our environment as Zurek and Paz 
\cite{Zurek2}. The spectral
density is Ohmic but a high temperature environment is also required. 
Consequently, the parameters are chosen so that $k_{B} T > \hbar \omega_{C}$.

\subsection{Qualitative Comparisons}

\subsubsection{Local KS Entropy}

When the nonlinearity parameter, $K$, of our rotor is greater than 
or equal to 6 there are no discernable stable islands in phase space plots 
of the classical standard map. They do still exist, of course,
but the fraction of the unit square they occupy at these values is small.
In such a situation the variation in the Lyapunov exponent will be 
small throughout phase space and the inverted oscillator model will be 
more likely to be considered adequate.
However, when the value of $K$ is so small as to allow KAM islands to exist 
and occupy a significant fraction of phase space we must expect the 
{\em local} Lyapunov exponents to vary considerably. 
All of these local Lyapunov exponents contribute to the sum in the 
calculation of the {\em global} KS entropy in equation (\ref{ModCon}). 
However, we can also define a {\em local} KS entropy by {\em restricting} 
the sum over local Lyapunov exponents to those corresponding to the points in 
an initial classical probability distribution.    
Thus, in view of the aim of this investigation, we might expect that the rate
of von Neumann entropy increase will depend strongly upon precisely where in 
a mixed space we locate an initial coherent state, i.e. upon the local 
KS entropy corresponding to its classically analogous initial probability
distribution. 

To illustrate the usefulness of the local KS entropy concept we considered 
first of all the time evolution of two initially 
Gaussian, symmetric probability distributions in phase space, one centred in 
a classically regular region when $K = 1$, with 
the other centred in a chaotic region for the same $K$ value. The variance of 
these distributions was chosen as $\hbar / 2$ in order to render them classical
analogues of the minimum uncertainty quantum initial states to which their
evolutions will be compared.

\begin{figure}[h]
\epsfxsize=6in
\epsfysize=6in
\epsffile{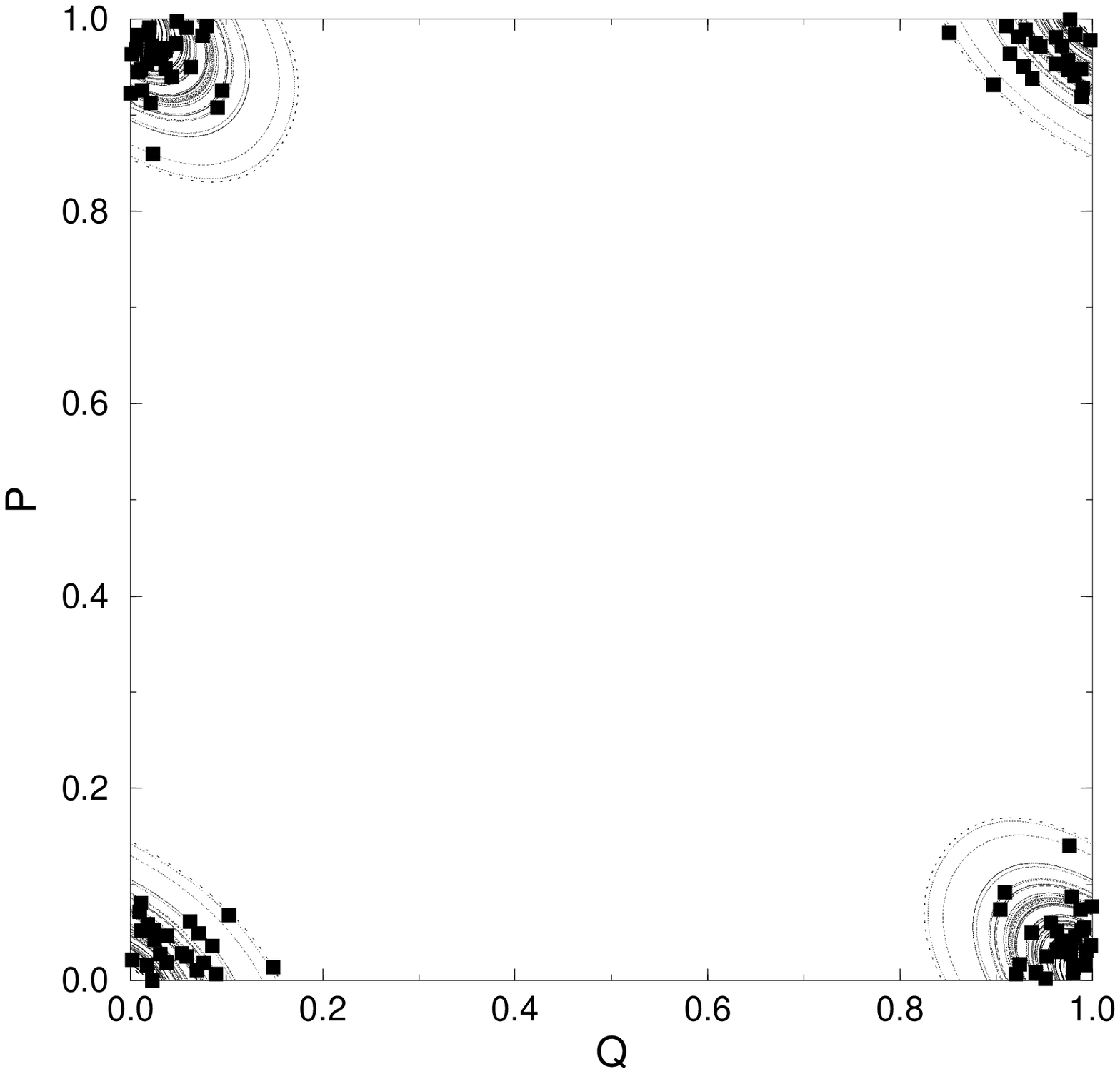}
\caption{The squares are 100 sample points from an initial Gaussian 
distribution centred at the origin $(0.0, 0h)$ with a variance of $\hbar / 2$ 
corresponding to a quantum initial state of minimum uncertainty. We choose 
$\hbar = 0.05 \gamma / (2 \pi)$. The dots mark the first 300 iterations of 
each such initial point, thereby giving an indication of the predominantly 
regular resultant trajectories.}
\label{fig1}
\end{figure}

\begin{figure}[h]
\epsfxsize=6in
\epsfysize=6in
\epsffile{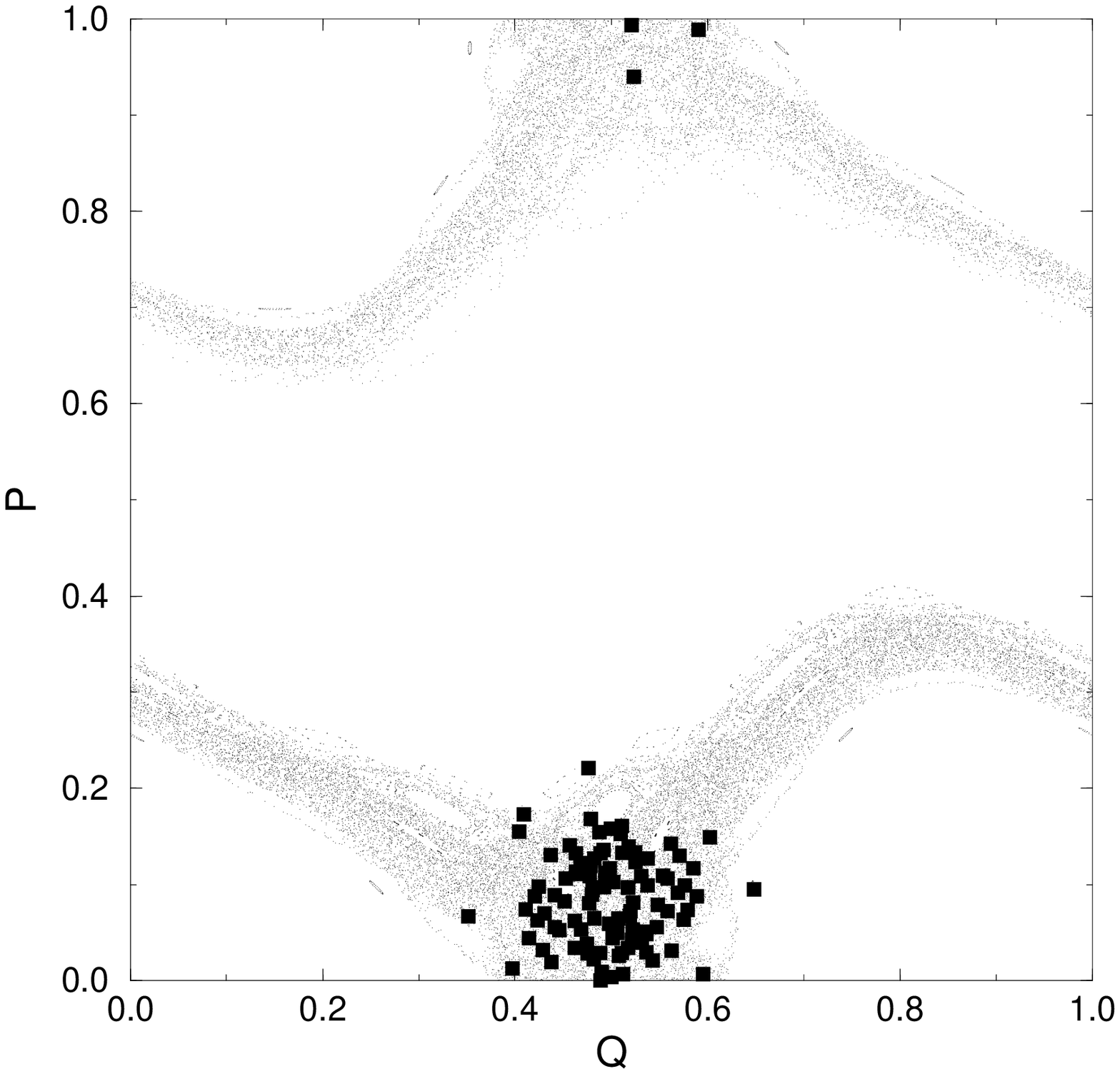}
\caption{The squares are 100 sample points from an initial Gaussian 
distribution centred at $(0.5, 2h)$ with a variance of $\hbar / 2$ 
corresponding to a quantum initial state of minimum uncertainty. We choose 
$\hbar = 0.05 \gamma / (2 \pi)$. The dots mark the first 300 iterations of 
each such initial point, thereby giving an indication of the predominantly 
chaotic resultant trajectories.}
\label{fig2}
\end{figure}

In figures [\ref{fig1}] and [\ref{fig2}] we plot the result of taking 100 
sample points from each of these distributions and plotting the first 300
points of the classical trajectories corresponding to each such point. The
regularity of the dynamics in figure [\ref{fig1}], where the initial state 
is centred at
$(0.0, 0h)$, $h$ being Planck's constant as before, can be compared to the
chaotic trajectories in figure [\ref{fig2}] resulting from the initial state 
being 
centred at $(0.5, 2h)$. 

Our next step is to calculate the KS entropy \cite{Chirik} corresponding to 
each of the points 
used to sample the Gaussians and then to take the average. This we define as
the local KS entropy corresponding to the initial states. 
The calculations are done for a range of $K$ values from $0$ to $6$ giving 
rise to progressively less mixed phase spaces.
In figure 
[\ref{fig3}] we see that this {\em local} KS entropy differs greatly 
for these two states when $1 \leq K \leq 4$, i.e. after a significant number of
invariant KAM curves have been destroyed and before the $K$ value at which 
the KAM island surrounding the elliptic fixed point at the origin disappears 
completely from phase space plots \cite{Reichl,Chirik}. As $K$ increases 
further these local KS entropies agree more and more, and can be well 
approximated by the {\it global} KS entropy \cite{Chirik} of equation 
(\ref{KSent}).

\subsubsection{Local Quantum-Classical Entropy Correspondence} 

Though global KS entropy can, of course, be defined for {\em all} values of 
$K$, it must be numerically determined when a largely mixed phase space 
prevails. We show here, however, that it is not the quantity to be used in 
any prediction of the quantum mechanical entropy production rate. The correct 
quantity is the local KS entropy. 

\begin{figure}[h]
\epsfxsize=6in
\epsfysize=6in
\epsffile{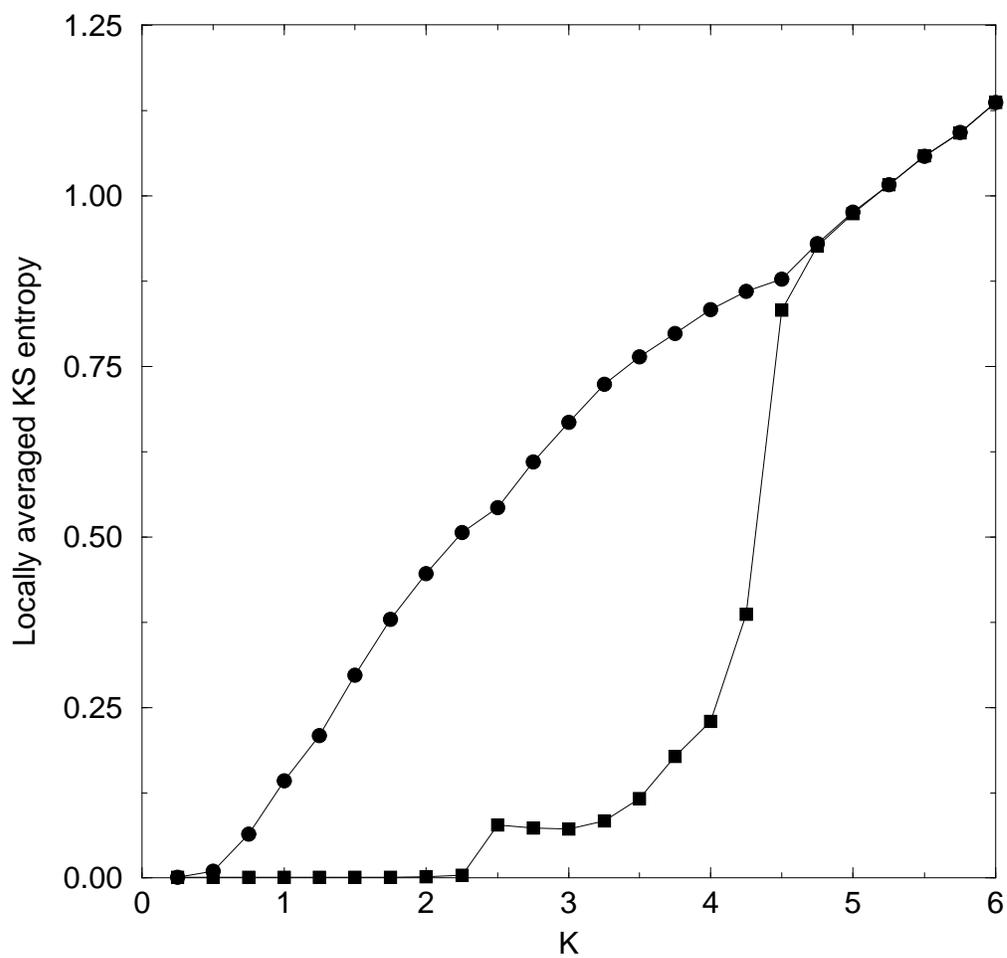}
\caption{The locally averaged KS entropy of each distribution in figures
[\ref{fig1}] and [\ref{fig2}], corresponding to the filled squares and the
filled circles, respectively, plotted against $K$.}
\label{fig3}
\end{figure}

\begin{figure}[h]
\epsfxsize=6in
\epsfysize=6in
\epsffile{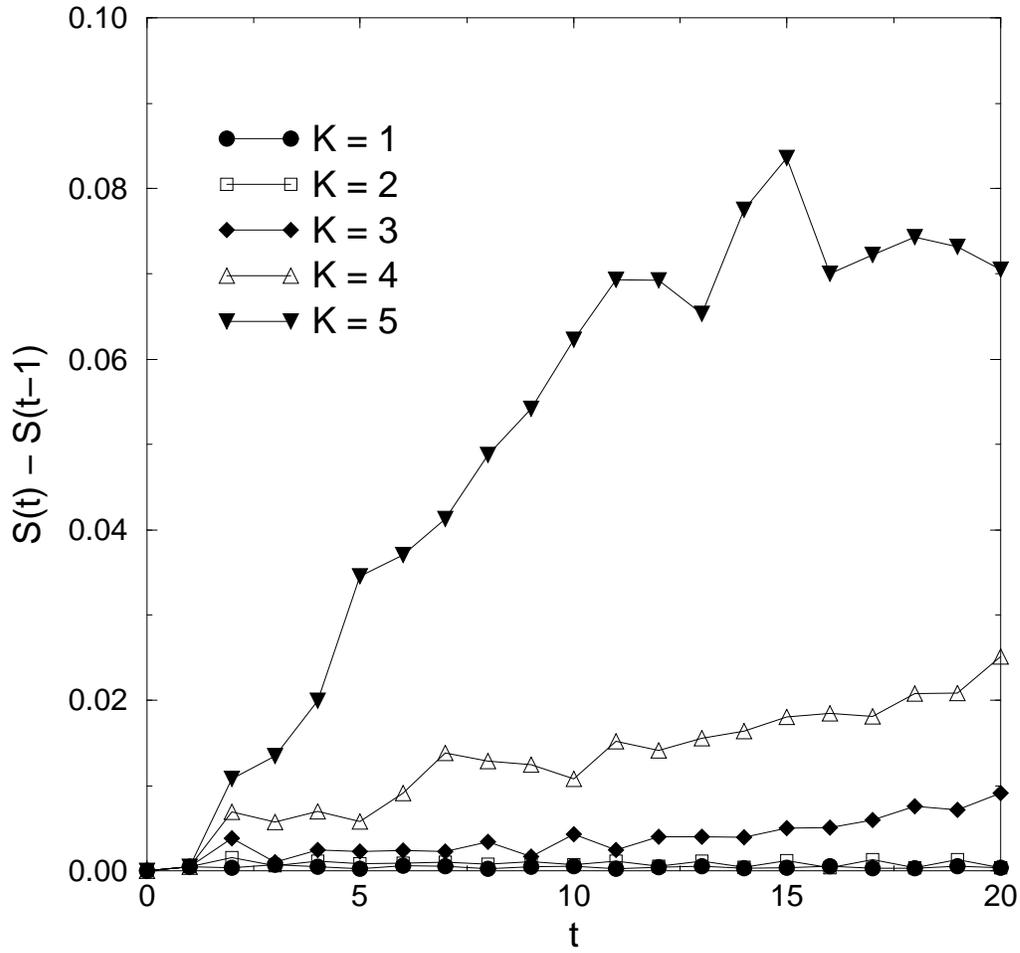}
\caption{The change in the entropy production rate, $h_{Q}$, with time, $t$,
 for various values of $K$. The initial state corresponds to that 
plotted in figure [\ref{fig1}], giving rise to regular trajectories. 
Parameter values used are $h = 0.1 \gamma, \beta = 0.1$.}
\label{fig4}
\end{figure}

\begin{figure}[h]
\epsfxsize=6in
\epsfysize=6in
\epsffile{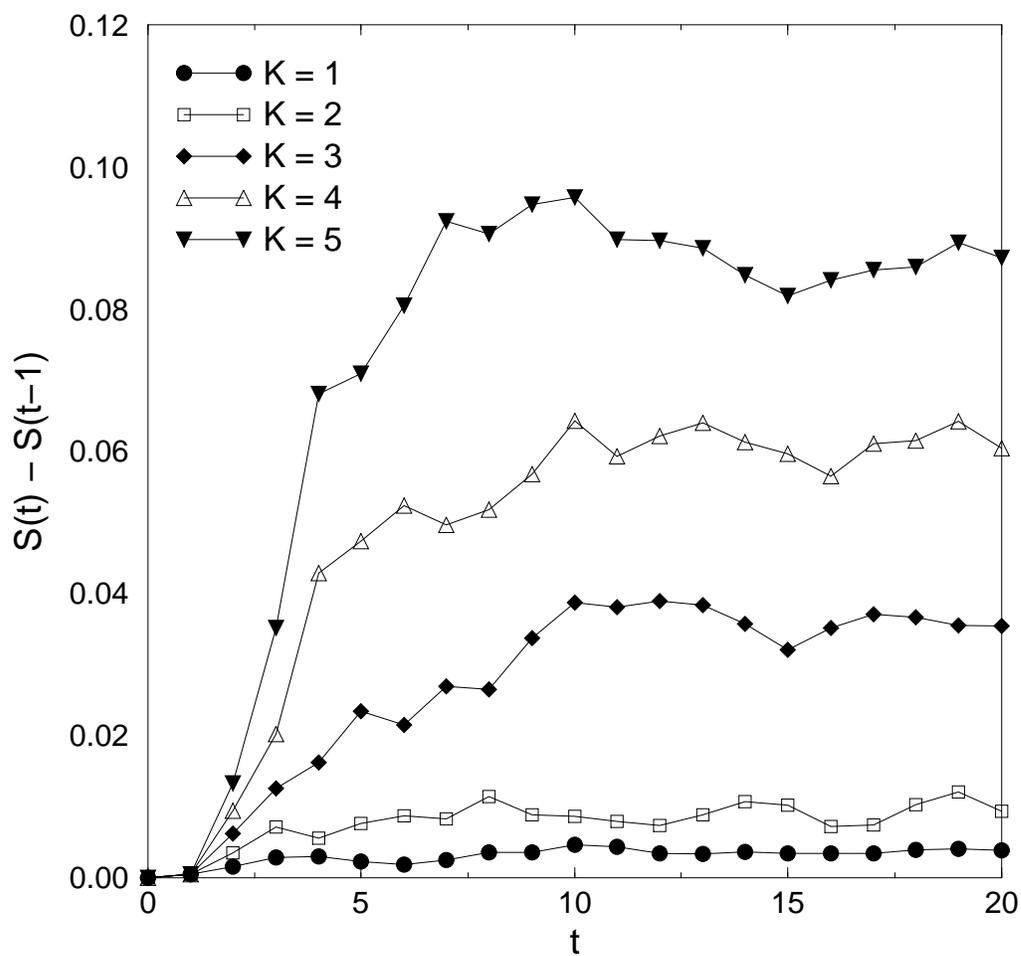}
\caption{The change in the entropy production rate, $h_{Q}$, with time, $t$,
 for various values of $K$, and using the same parameter values of 
figure [\ref{fig4}]. The initial state corresponds to that 
plotted in figure [\ref{fig2}], giving rise to chaotic trajectories.}
\label{fig5}
\end{figure}

To show this we next considered the quantum mechanical time evolution 
of the two coherent states analagous to the two classical distibutions used 
above. For each state we have calculated the von Neumann entropy production
rate over 20 time steps and for the set $K = 1, 2, \ldots, 5$, values at 
which the locally calculated KS entropies differ greatly 
(See figure [\ref{fig3}]). 
The results corresponding to the initial, classically regular state centred 
at $(0.0, 0h)$ can be seen in figure [\ref{fig4}] and 
those corresponding to the (classically chaotic) state centred at $(0.5, 2h)$ 
can be seen in figure [\ref{fig5}]. In the former case the entropy 
production rates reflect very well 
the local KS entropy values of figure [\ref{fig3}]. The  initially low values 
of each, when $K =
1, 2$, increase when $K = 3$, once more when $K = 4$, and finally jump
dramatically when $K = 5$. However, the situation for the coherent state
situated initially in a classically chaotic region of phase space is 
strikingly different. Indeed, the steady increase in plateau height with 
$K$ mirrors well the local KS entropy of figure [\ref{fig3}] as it, too, 
increases steadily
with $K$. No sudden jumps are apparent, and only when $K = 5$ do the entropy 
production rates of both initial states approach one another, just as they do
classically.

\subsubsection{Overlap Effects in a Mixed Phase Space} 

\begin{figure}[h]
\epsfxsize=6in
\epsfysize=6in
\epsffile{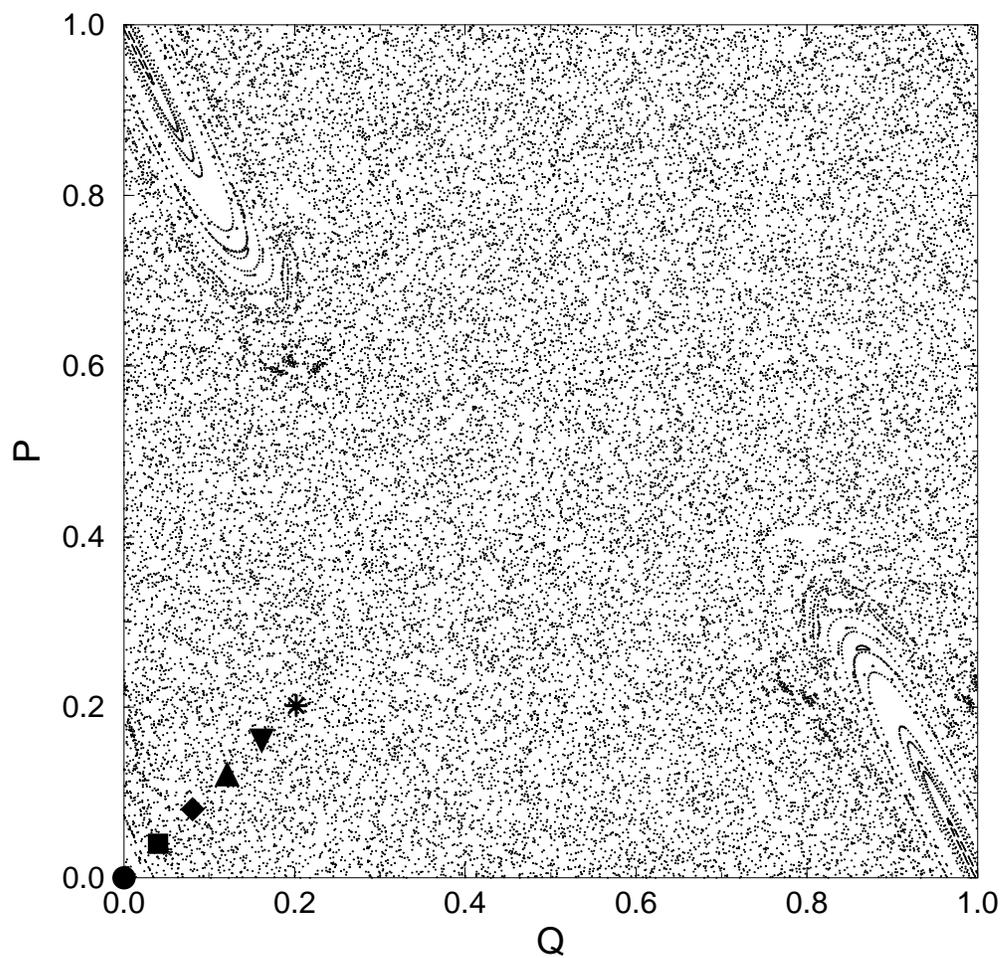}
\caption{A classical phase space plot for $K = 4$ with the stable island 
clearly visible. The superimposed symbols correspond to those in 
figure [\ref{fig7}] and mark the centre of each initial coherent state for
which the entropy rate is plotted in that figure.}
\label{fig6}
\end{figure}

\begin{figure}[h]
\epsfxsize=6in
\epsfysize=6in
\epsffile{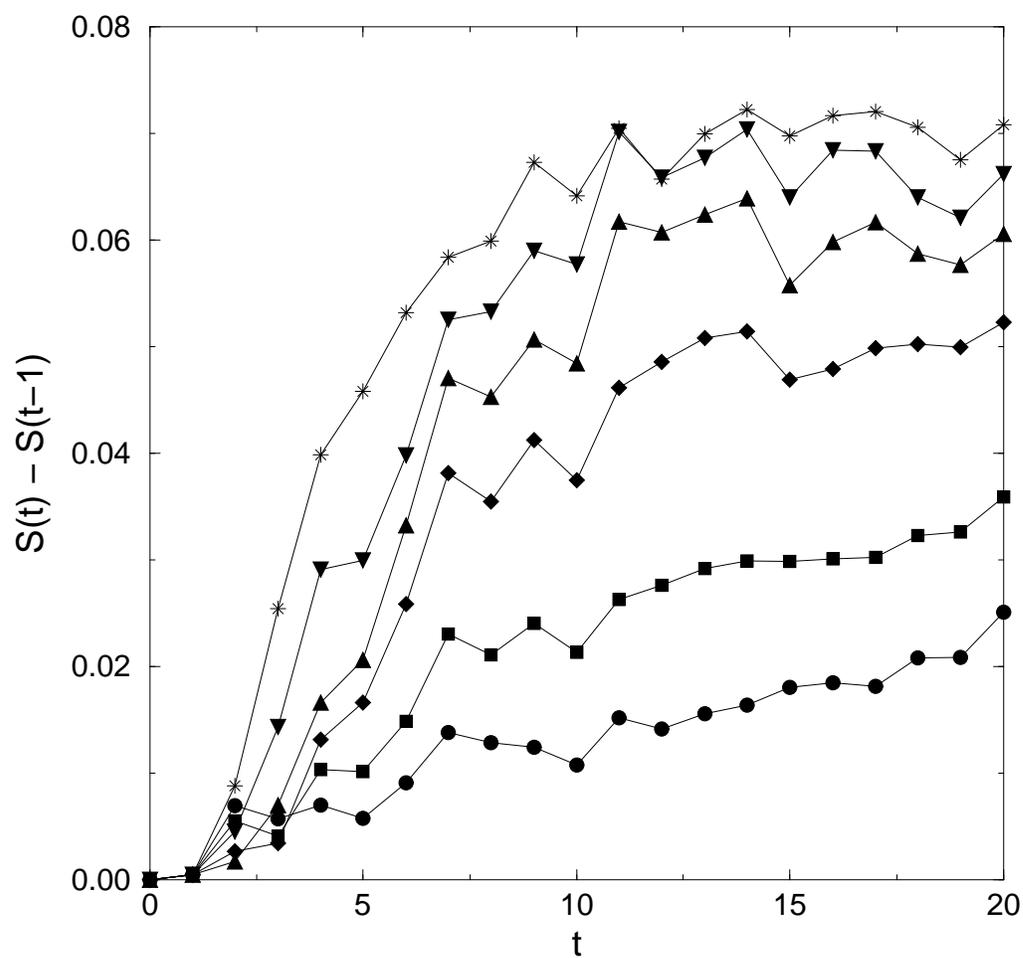}
\caption{The rate of von Neumann entropy production for various starting 
points in the mixed phase space of figure [\ref{fig6}] and denoted by the same
symbols here. Here, $h = 0.05 \gamma$.}
\label{fig7}
\end{figure}

\begin{figure}[h]
\epsfxsize=6in
\epsfysize=6in
\epsffile{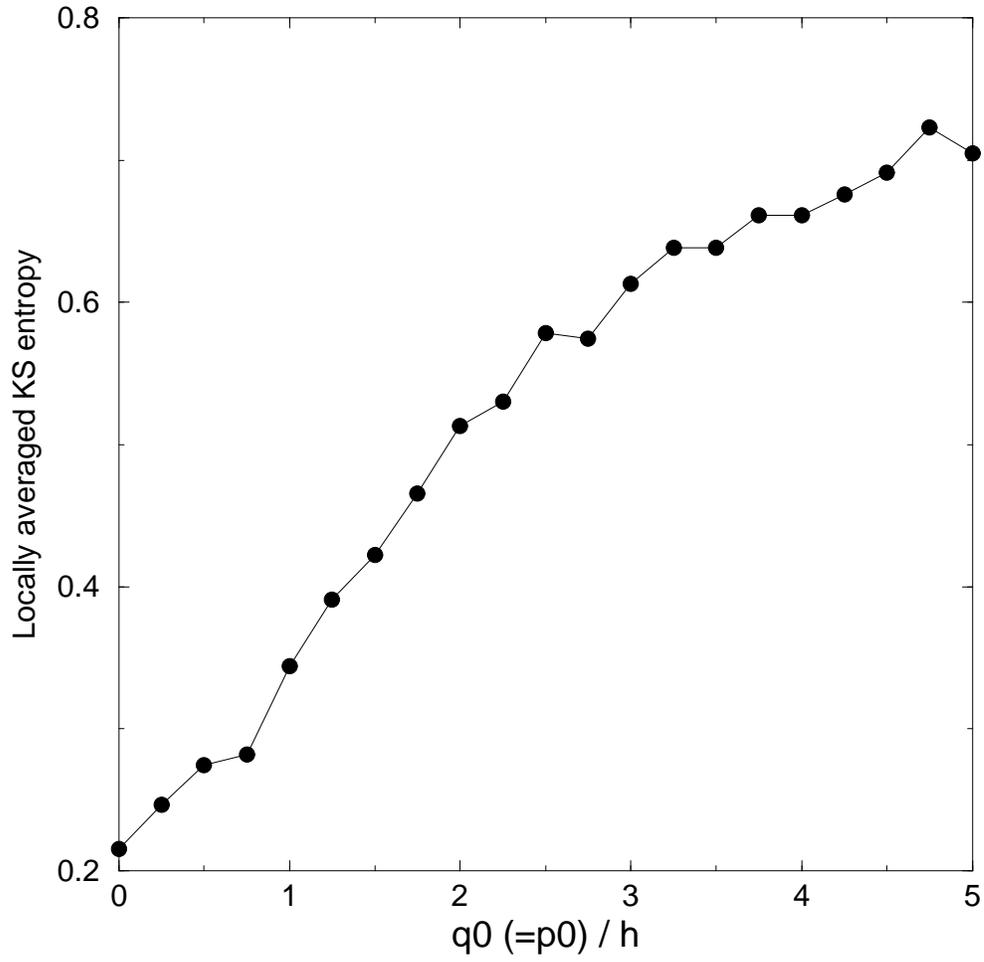}
\caption{The classically calculated KS entropy for various initial Gaussian 
distributions centred along the diagonal $q0 = p0$ in the mixed phase space of 
figure [\ref{fig6}]. The position of the filled circles along the horizontal
axis indicates the centre of the initial distribution. Again we choose 
$h = 0.05 \gamma$.}
\label{fig8}
\end{figure}

Entropy production, thus, {\em does} depend on where we initially place our
coherent state but we expect this variability to decrease with increasing 
values of $K$ as the classical phase space becomes more uniformly chaotic. 
However, the initial states for $K = 1$ we have considered above
have been well localized (figures [\ref{fig1}] and [\ref{fig2}]) 
in either entirely regular or entirely chaotic regions, with no significant 
overlap with nearby regions which may have very different classical behaviour.
In order to investigate the effect significant overlaps have on the rate of 
entropy 
production we set $K = 4$, where, as can be seen from the classical phase 
space plot of figure [\ref{fig6}], one large regular region still exists in 
a sea of chaotic trajectories. If we    
locate our initial coherent states along the diagonal which intercepts this 
KAM island, with
their centres equally spaced at $(0,0), (h,h), (2h,2h), \ldots$, we find an 
interesting feature. The rate of entropy increase is lower when the 
initial state overlaps the regular region and only when a significant 
overlap exists initially with the chaotic sea does the entropy begin to 
increase at the robust rate which is found for all the states with little or 
no overlap with this island. Classically we see analagous behaviour once more 
if we examine the locally averaged KS entropy for initial distributions which
correspond in their mean and variance to those quantum states of figure 
[\ref{fig7}]. We show the results of this calculation in figure [\ref{fig8}]. 
Here, as in the quantum case, an asymptotic approach to a constant KS entropy 
value (of $\approx0.8$) is found as the initial distibution overlaps to a 
lesser degree with the stable KAM island.     

It seems, therefore, that entropy production is indeed a suitable criterion
by which to distinguish between regular and chaotic motion. Recently Zarum 
and Sarkar \cite{Zarum1} produced qualitative evidence in support
of this claim for the kicked rotor model. They quantized on a torus and 
perturbed the rotor in a different, but noisy and non-unitary way, leading to 
an entropy 
increase. Initial coherent states uniformly distributed on the torus gave rise 
to a corresponding distribution of (constant) entropy production rates which
reflect in a remarkable way the classical KS entropy profile for initial
points in a mixed phase space. Similar results have recently been obtained 
for the quantum kicked top model \cite{Zarum2}.

\subsection{Quantitative Comparisons}

The {\em quantitative} analysis of the relation between $h_{Q}$ and 
$h_{KS}$ has been the goal of this paper and we now turn our attention to it.

\subsubsection{Temperature-induced Correspondence}

\begin{figure}[h]
\epsfxsize=6in
\epsfysize=6in
\epsffile{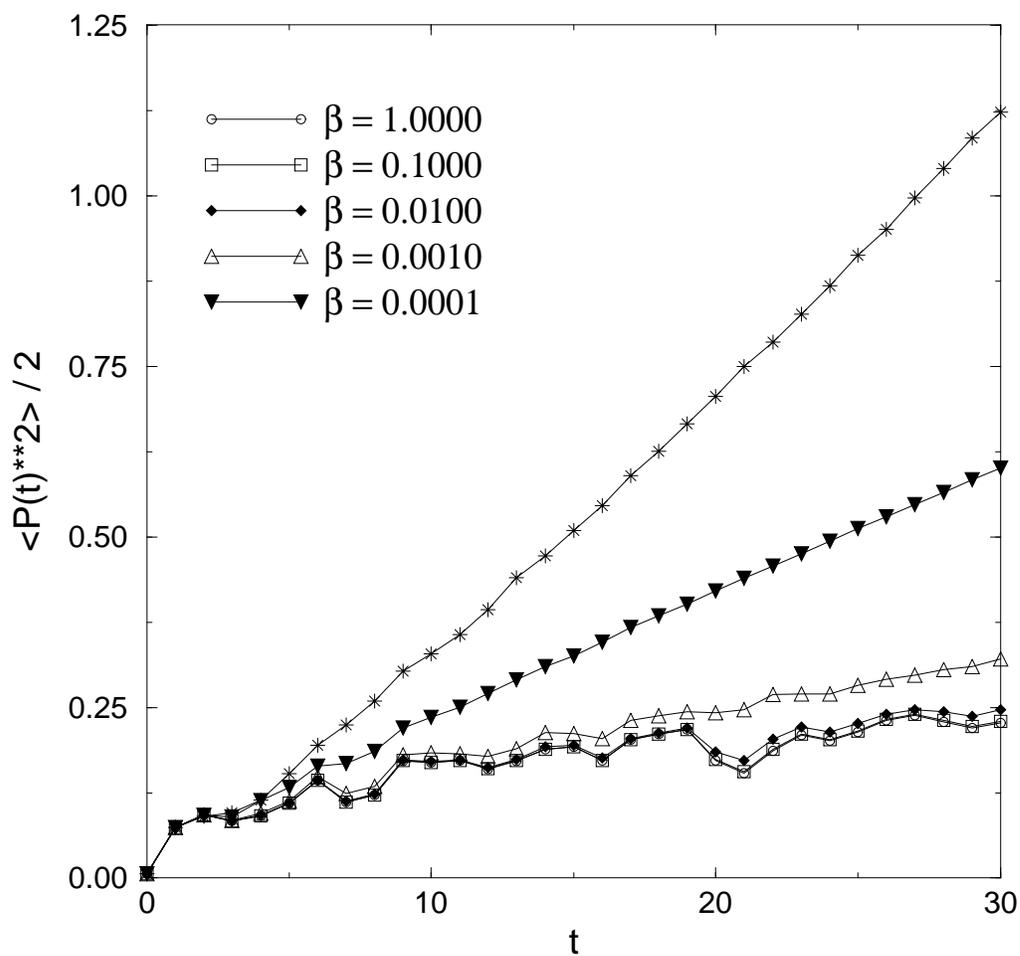}
\caption{Here we see that increasing the temperature of the environment
(equivalently, decreasing $\beta$) has the effect of restoring the
correspondence between quantum and classical expectation values. See the text 
for details.}
\label{fig9}
\end{figure}

Firstly we examine the effect that increasing the environmental temperature 
has on quantum-classical correspondence. To do so we have once more iterated 
in time a normally distributed initial classical state consisting of 
25000 points. The distribution was chosen to be centred at the same point, and 
to have the same variance, as a minimum uncertainty quantum state centred at 
$(0.5, h)$. We have chosen $h = 0.1 \gamma$ and $K = 3$ to examine the 
correspondence in time between the classical average value and the quantum 
expectation value of the squared momentum variable in figure [\ref{fig9}]. 
Clearly, the classical behaviour, marked with stars ($\ast$), is approached 
to a greater degree as the temperature is increased (equivalently, as 
$\beta = 1/k_{B}T$ is decreased) despite the fact that $\hbar$ {\it remains 
fixed}. 

\begin{figure}[h]
\epsfxsize=6in
\epsfysize=6in
\epsffile{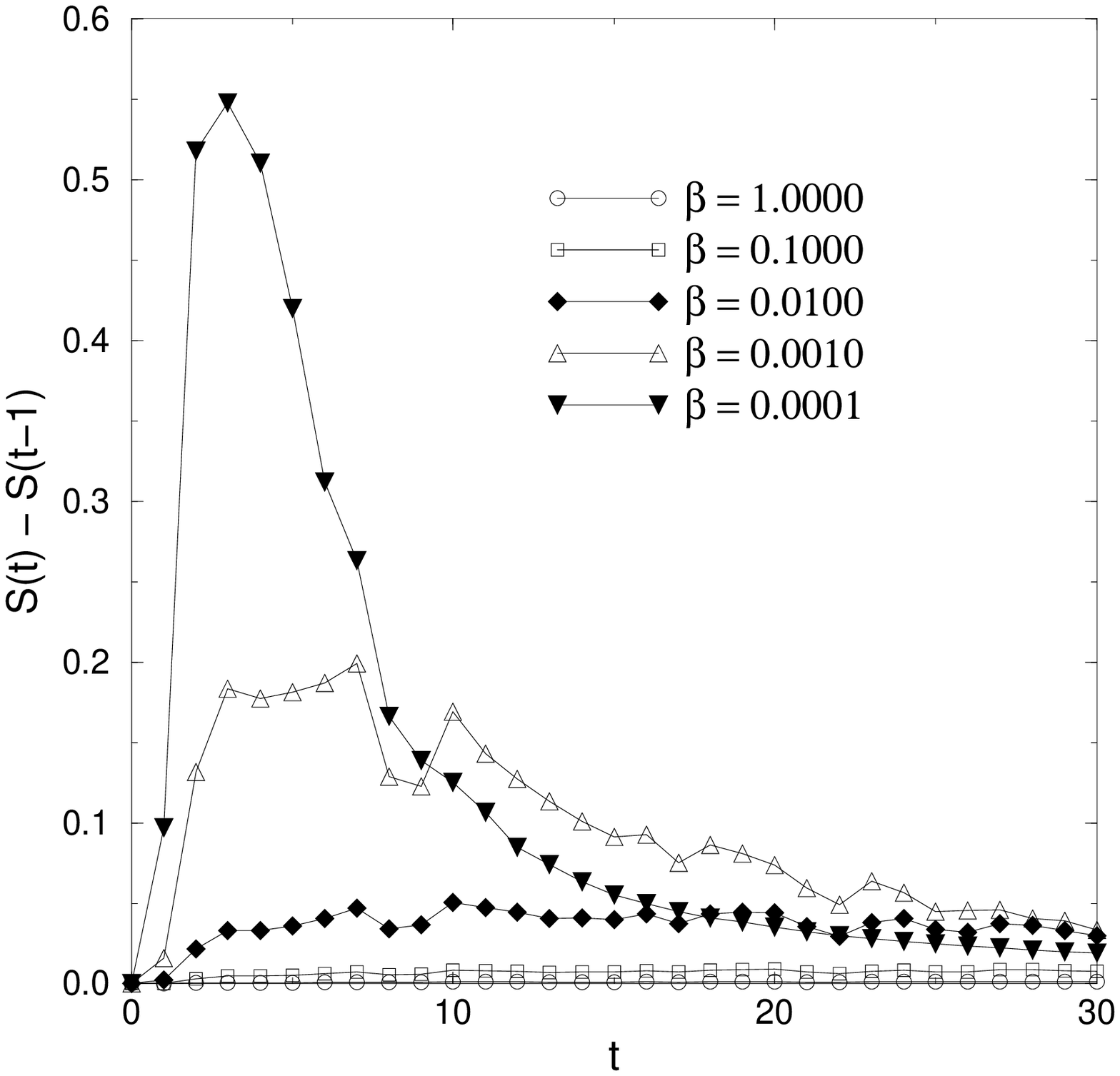}
\caption{Here we see the consequences that restoration of the 
quantum-classical correspondence has for the von Neumann entropy production 
rate. See the text for details.}
\label{fig10}
\end{figure}

\begin{figure}[h]
\epsfxsize=6in
\epsfysize=6in
\epsffile{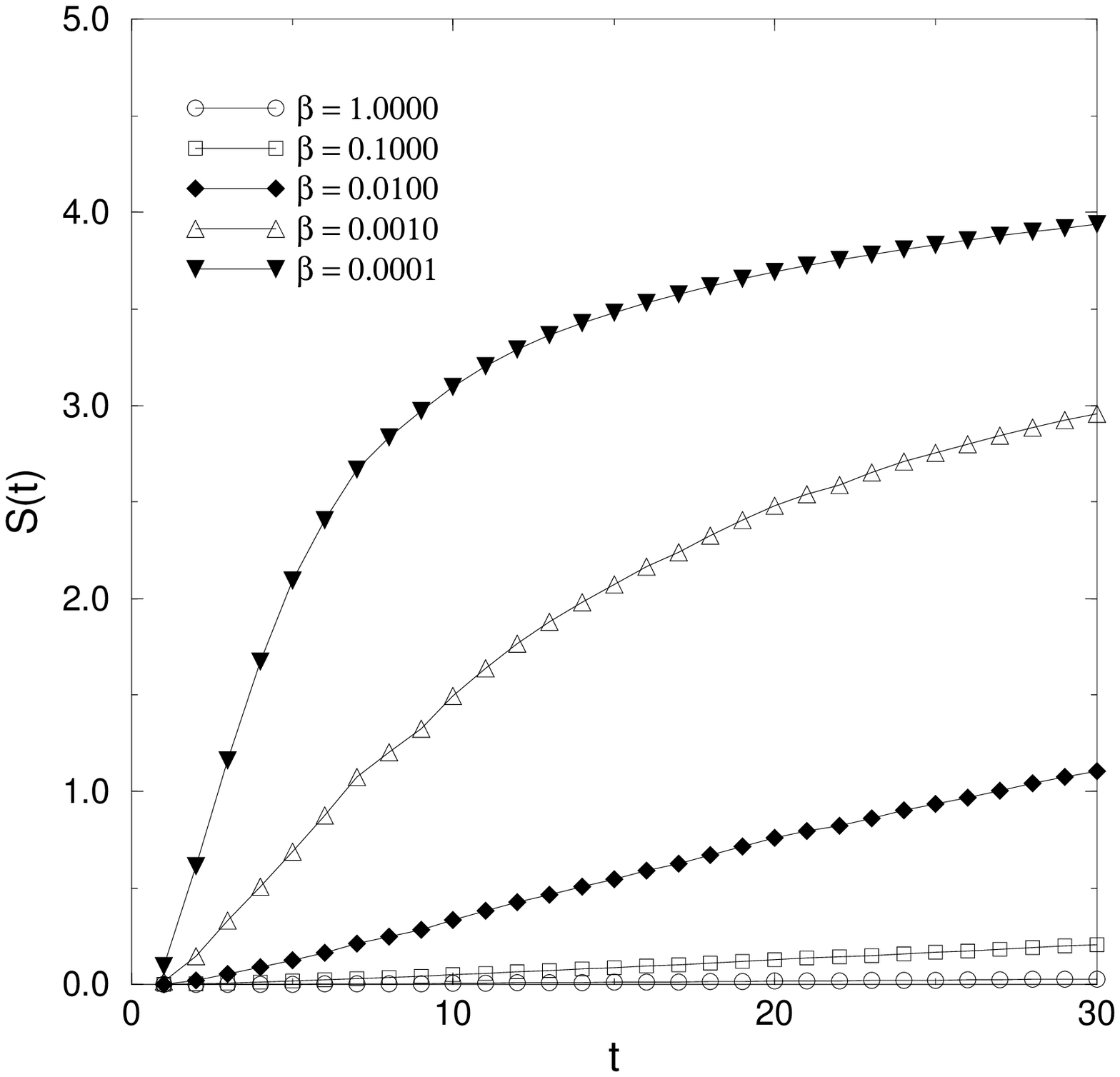}
\caption{Here we see the consequences that restoration of the 
quantum-classical correspondence has for the von Neumann entropy. 
See the text for details.}
\label{fig11}
\end{figure}

Now, in figure [\ref{fig10}] we examine, for the same parameters, the 
consequences for the rate of von Neumann entropy production and in 
figure [\ref{fig11}] the consequences for the von Neumann entropy itself. The
truncated basis chosen for these runs as $\{|-n \rangle, |-n+1 \rangle , 
\ldots, |-1 \rangle, |0 \rangle, |+1 \rangle, \ldots, |n-1 \rangle, 
|n \rangle\}$, where $n = 250$. This implies a maximum entropy value of 
$\ln 501 \approx 6.2$. 

For inverse temperatures $\beta = 1.00, 0.10$ and 
$0.01$, and for the time regime shown, we find that the von Neumann entropy 
is not near enough to the saturation value for the linearity of its increase 
with time to be affected. Eventually, however, the entropy will begin to
saturate even for these low temperatures and the rates of entropy increase will 
fall below that of their respective ``plateaux" shown in figure [\ref{fig10}]. 
A feature {\em not} expected on the basis of the inverted oscillator model 
\cite{Zurek2,paper2}, and yet clearly in evidence here, is the dependence on 
temperature of the plateau height. Hence the need for a {\em modified} 
conjecture to take this effect into account.  

When we increase the temperature still further, so that $\beta = 0.001$, we 
note two generic features. Firstly, the time at which the approximately 
constant entropy production is attained, i.e. the time at which the plateau
begins, decreases with the increase in temperature. This is due to the greater 
effectiveness of high-temperature environments in destroying quantum 
coherence. Secondly, we note that the plateau exists for a shorter length of 
time as the temperature is increased. This is because the increase in 
the plateau height due to the increase in temperature causes the entropy to 
approach its saturation value at an earlier time (by $t = 8$ here). 
We note that the first of these features was mentioned by Zurek and Paz in the 
framing of their original conjecture \cite{Zurek2}. They also expected a 
finite lifetime for the plateaux due to saturation effects. However, the 
increase of plateau height with temperature seen here shortens this lifetime 
still further, making the plateaux seem more and more like maxima as the 
temperature is increased.
 
For the highest temperature chosen, with $\beta = 0.0001$, we see 
that the initial coherence is lost so quickly as to render the dynamical 
localization mechanism ineffective. Indeed, a linear increase of 
$\langle P^{2}(t) \rangle / 2$ with the time, $t$, is seen in 
figure [\ref{fig9}] at this temperature, albeit at a lower rate than the 
classical value. This discrepancy, however, disappears as the temperature is
increased still further. With regard to entropy and its production rate we see 
the above-mentioned features strikingly displayed: the entropy production 
rate has a maximum rather than a plateau; this maximum is greater than the 
plateaux heights of lower temperatures, and the maximum is reached at the 
shorter time of $t = 3$.

\subsubsection{Factors Affecting the Decay Probability}

\begin{figure}[h]
\epsfxsize=6in
\epsfysize=6in
\epsffile{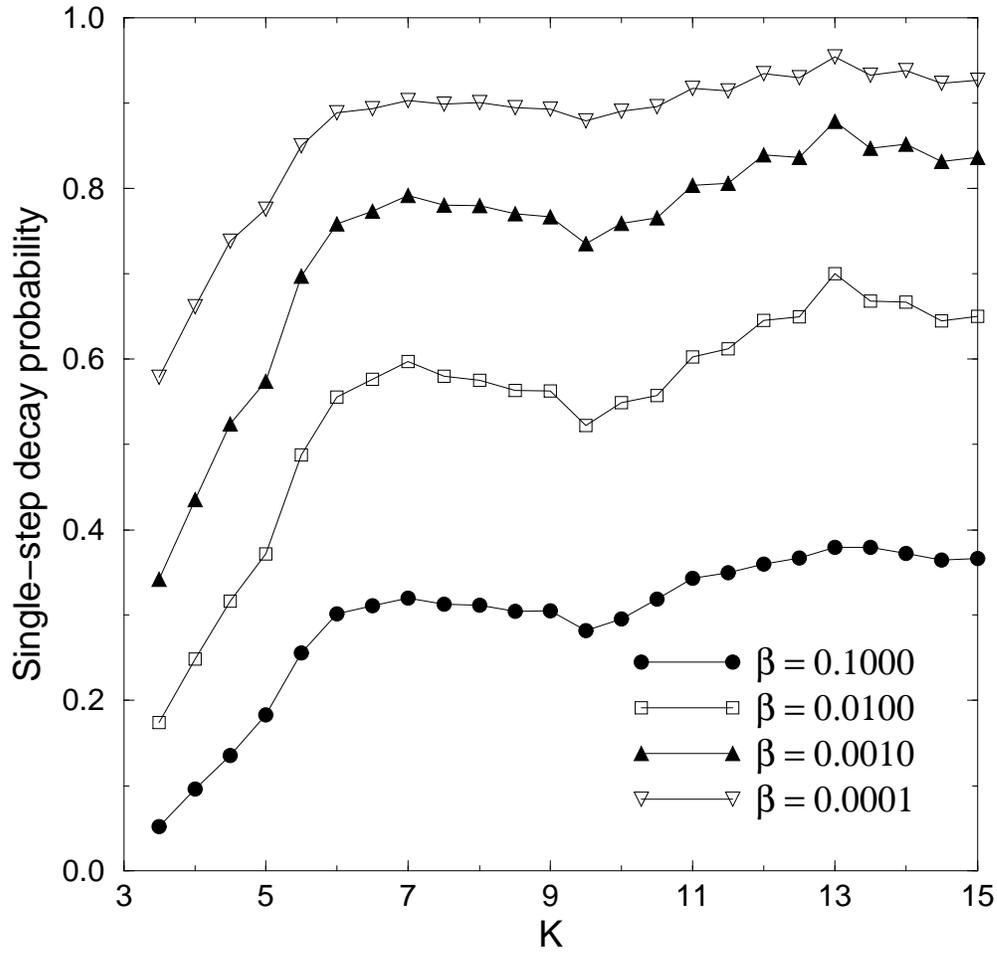}
\caption{In this figure we plot the variation of the single-step decay 
probability, $P_{1}$, with $K$, for various different temperatures. We have 
chosen $h = 0.1 \gamma$. See section \ref{section3} and section 
\ref{section5} for details.}
\label{fig12}
\end{figure}

\begin{figure}[h]
\epsfxsize=6in
\epsfysize=6in
\epsffile{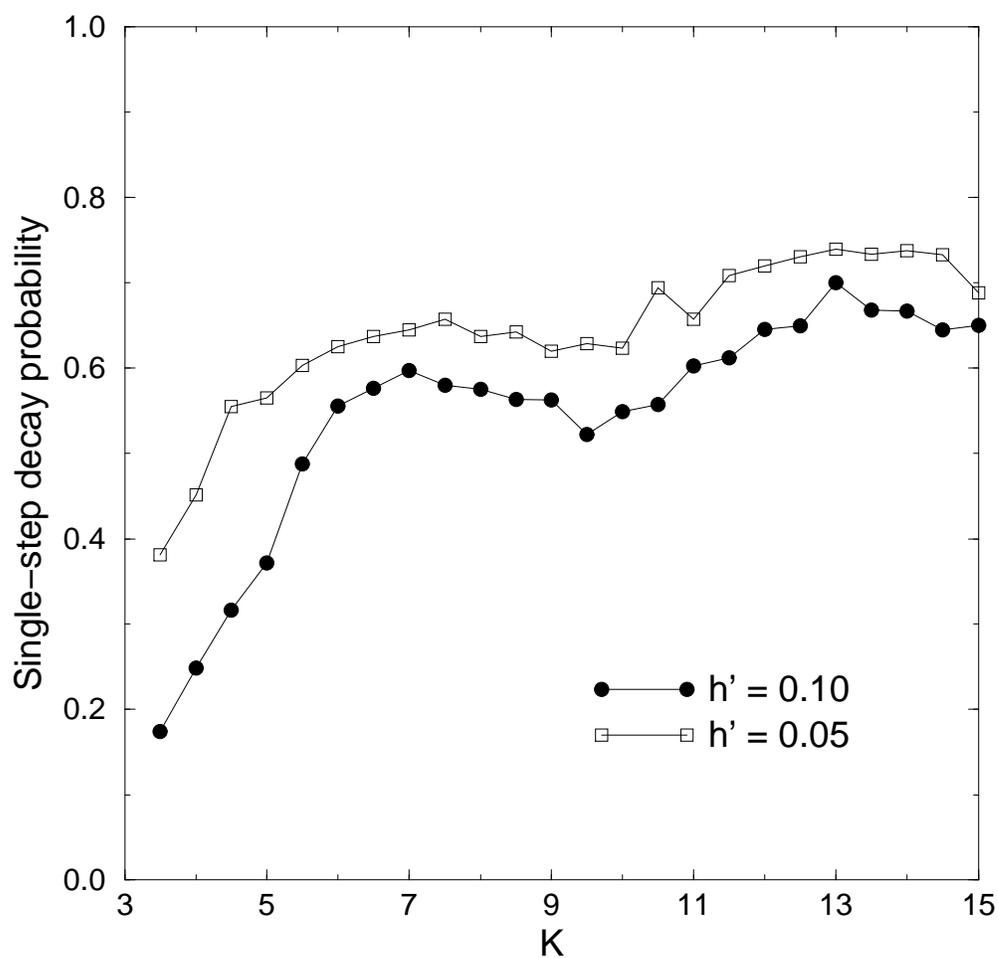}
\caption{In this figure we plot the variation of the single-step decay 
probability, $P_{1}$, with $K$, for two differnt values of $h$. We have 
chosen $\beta = 0.01$. See section \ref{section3} and section 
\ref{section5} for details.}
\label{fig13}
\end{figure}

Evidence that increasing one or more of the temperature, $h$ 
and $K$, the magnitude of the kick, destroys the quantum coherence more
efficiently can be seen in figures [\ref{fig12}] and [\ref{fig13}]. There 
we have plotted the single-step decay probability, $P_{1}$, described in 
section \ref{section3} against $K$, for different temperatures 
(figure [\ref{fig12}]) and values of the quantum of action 
(figure [\ref{fig13}]). These figures clearly indicate that to destroy the 
dynamical localization mechanism more efficiently one should change these 
parameters in this way in order to reduce the coherence time, $n_{c}$. 
Furthermore, by increasing $K$ and decreasing $h$ we also increase the 
value of the break time, $n^{*}$, according to equation (\ref{breaktime}). 
This gives the environment ``more time" to act as the situation 
$n^{*} \gg n_{c}$ develops. 

These observations are important in any
consideration of the rate at which the system produces entropy. 
Entropy production in the system is a manifestation of the increase in phase 
space area (or volume for higher dimensional systems) that a wave packet 
occupies when it is perturbed by an environment. 
Localization, therefore, and the consequent restrictions this places on the 
extent to which this area can increase, {\it must} be considered as a factor 
which can affect the entropy. We note that the inverted oscillator model, 
notwithstanding its usefulness as a guide to intuition, cannot take this 
effect into account.

\subsubsection{The Modified Conjecture}

\begin{figure}[h]
\epsfxsize=6in
\epsfysize=6in
\epsffile{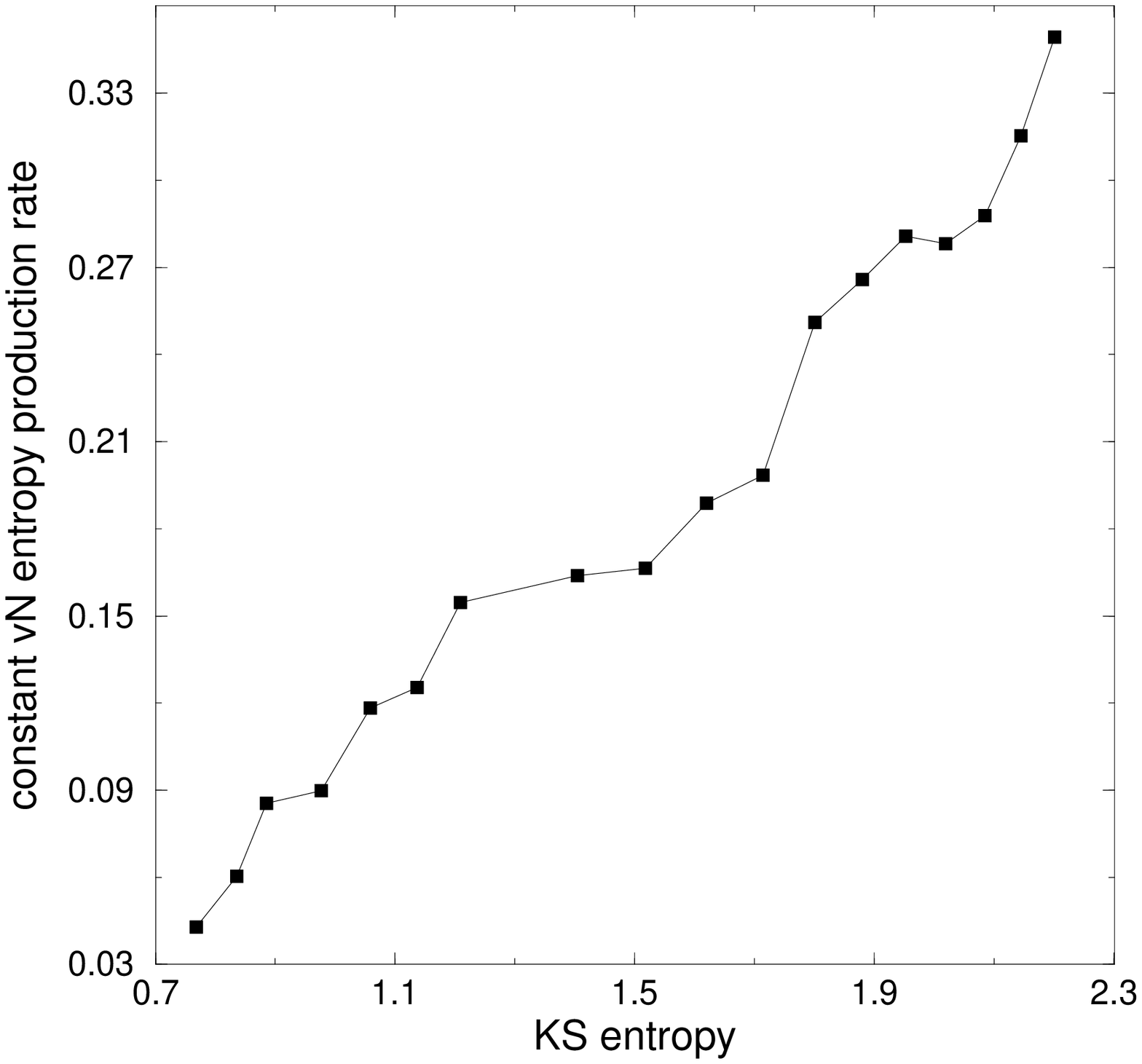}
\caption{The constant entropy production rate plotted against the 
corresponding local KS entropy. The starting point in 
space is $(0.5, 2h)$, where $h = 0.05 \gamma$. Here $\beta = 0.1$.}
\label{fig14}
\end{figure}

In order to examine the validity or otherwise of the modified conjecture we 
plot, in figures [\ref{fig14}] to [\ref{fig18}], the constant rate of 
von Neumann entropy production, $h_{Q}$, versus the corresponding, classical, 
local 
KS entropy for given initial states, with different environmental parameters 
and values of the quantum of action.

In figure [\ref{fig14}] we see a linear relationship but with {\it 
oscillations superimposed}. Here, $h = 0.05 \gamma$ and $\beta = 0.10$. 
Evidence that these oscillations have their 
origin in the dynamical localization mechanism can be gathered from the 
fact that the values of $K$ at which the plot departs most from linear 
behaviour are precisely those values for which the quasilinear approximation to
the classical diffusion coefficient, $D(K)$, is worst (see equation 
(\ref{DK}) and references \cite{LL,Rech}). Thus, on account of the link 
between both the localization length, 
$L$, and the break time, $n^{*}$, given by equations (\ref{loclength}) and 
(\ref{breaktime}), respectively, and the classical diffusion coefficient, 
we see that these values of $K$ give rise to exceptionally small or large 
localization lengths and break times. This, for the reason given above, 
affects the rate of entropy production. 
  
\begin{figure}[h]
\epsfxsize=6in
\epsfysize=6in
\epsffile{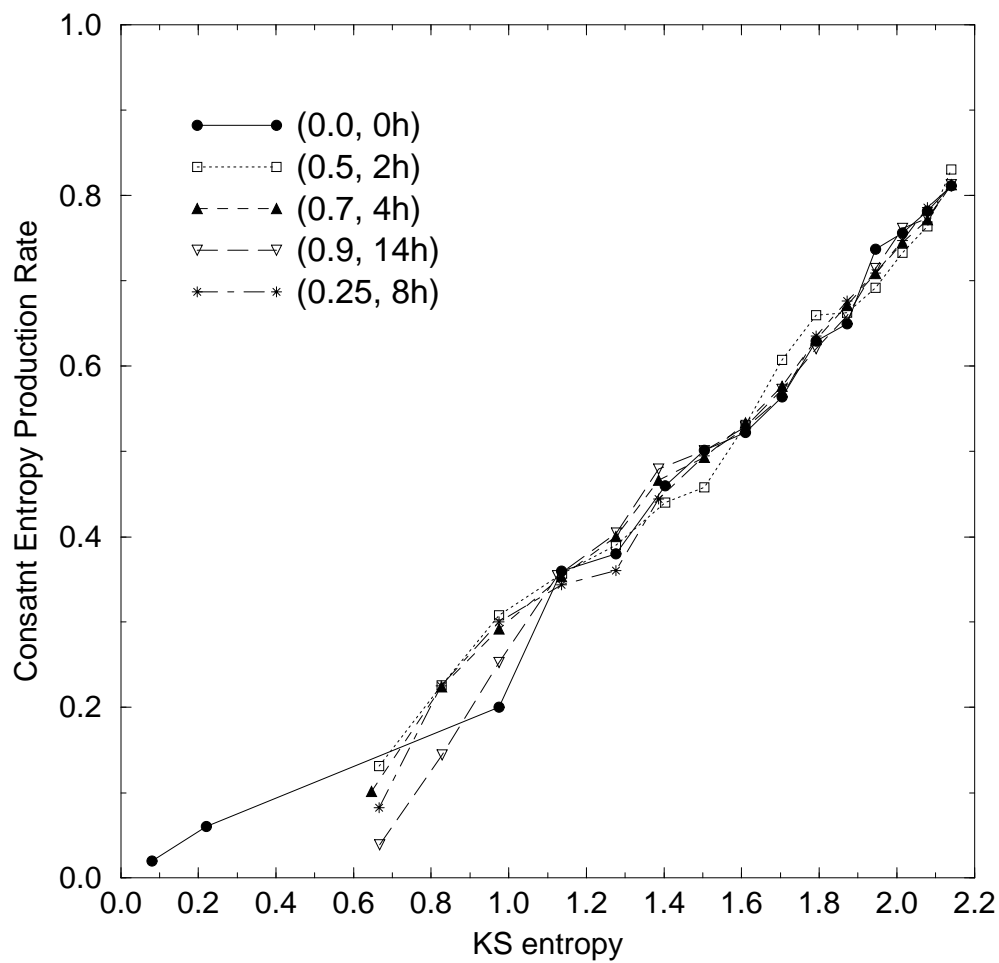}
\caption{The constant entropy production rate plotted against the 
corresponding local KS entropy for five different starting points in phase
space. Here, as in figure [\ref{fig12}] we choose $h = 0.05 \gamma$ but 
$\beta = 0.01$.}
\label{fig15}
\end{figure}

To investigate this effect further we have increased the temperature tenfold 
in figure [\ref{fig15}], i.e. $\beta = 0.01$. We have also chosen to start 
initial minimum uncertainty states at five different points in phase space. 
Since the quasienergy eigenstates form a complete set, each initial state 
can be expanded in this basis and will have different expansion coefficients. 
Differing levels of importance will then be attached to the various basis 
states for each initial condition. Since the localization length varies 
from eigenstate to eigenstate we might therefore expect to see slightly
different entropy production rates for each such state, but with the linear
behaviour still clearly in evidence. This is indeed the case. We  also note 
that the slope of the graph has increased with the increase in temperature.  

Running at higher temperatures when $h = 0.05 \gamma$ as in figures 
[\ref{fig14}] and [\ref{fig15}] quickly leads to undesirable saturation 
effects so we henceforth choose $h = 0.1 \gamma$ and examine the interplay 
between dynamical localization and the modified conjecture with this 
larger value for the quantum of action. 

\begin{figure}[h]
\epsfxsize=6in
\epsfysize=6in
\epsffile{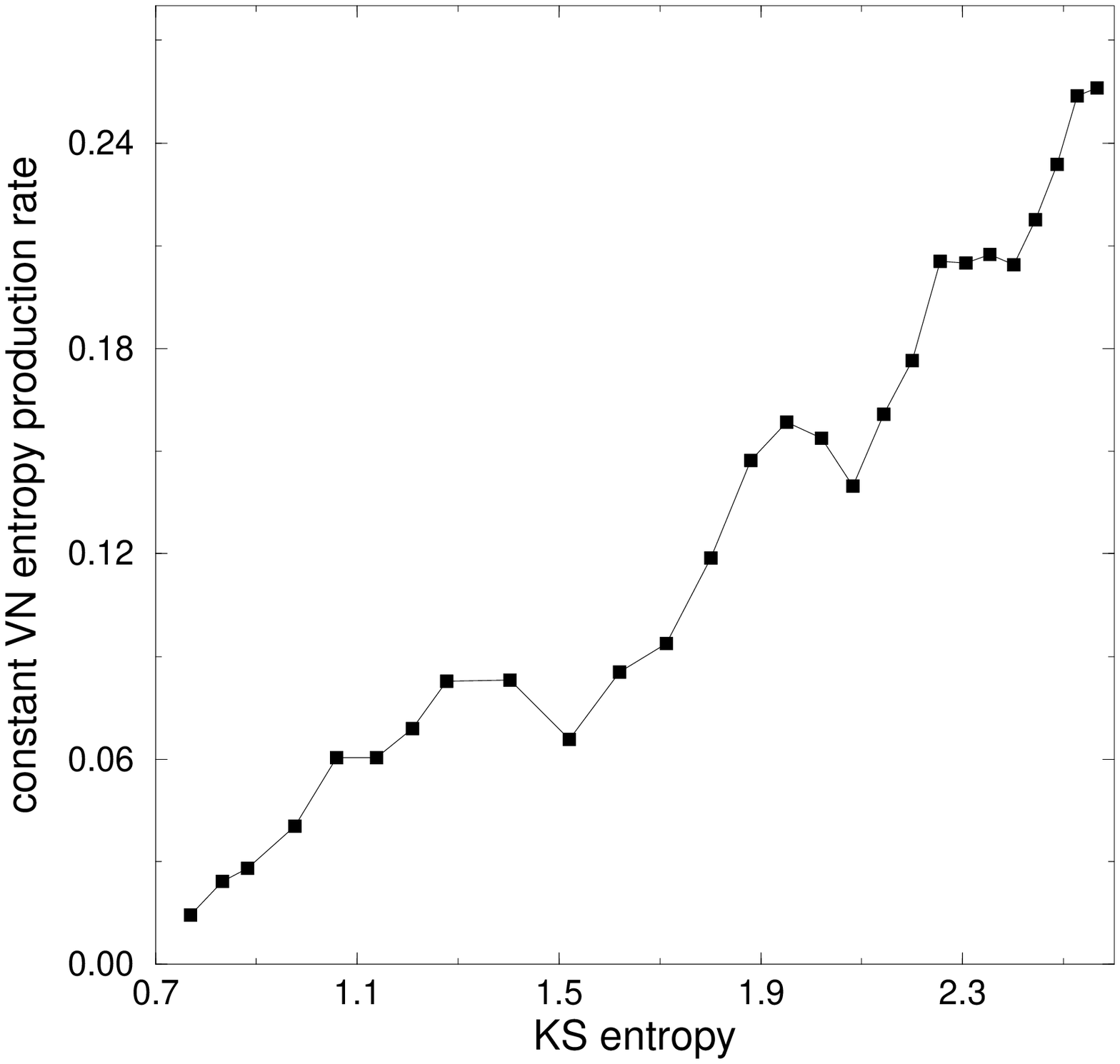}
\caption{The constant entropy production rate plotted against the 
corresponding local KS entropy. The starting point in phase 
space is $(0.5, h)$, where $h = 0.1 \gamma$. Here $\beta = 0.1$.}
\label{fig16}
\end{figure}

\begin{figure}[h]
\epsfxsize=6in
\epsfysize=6in
\epsffile{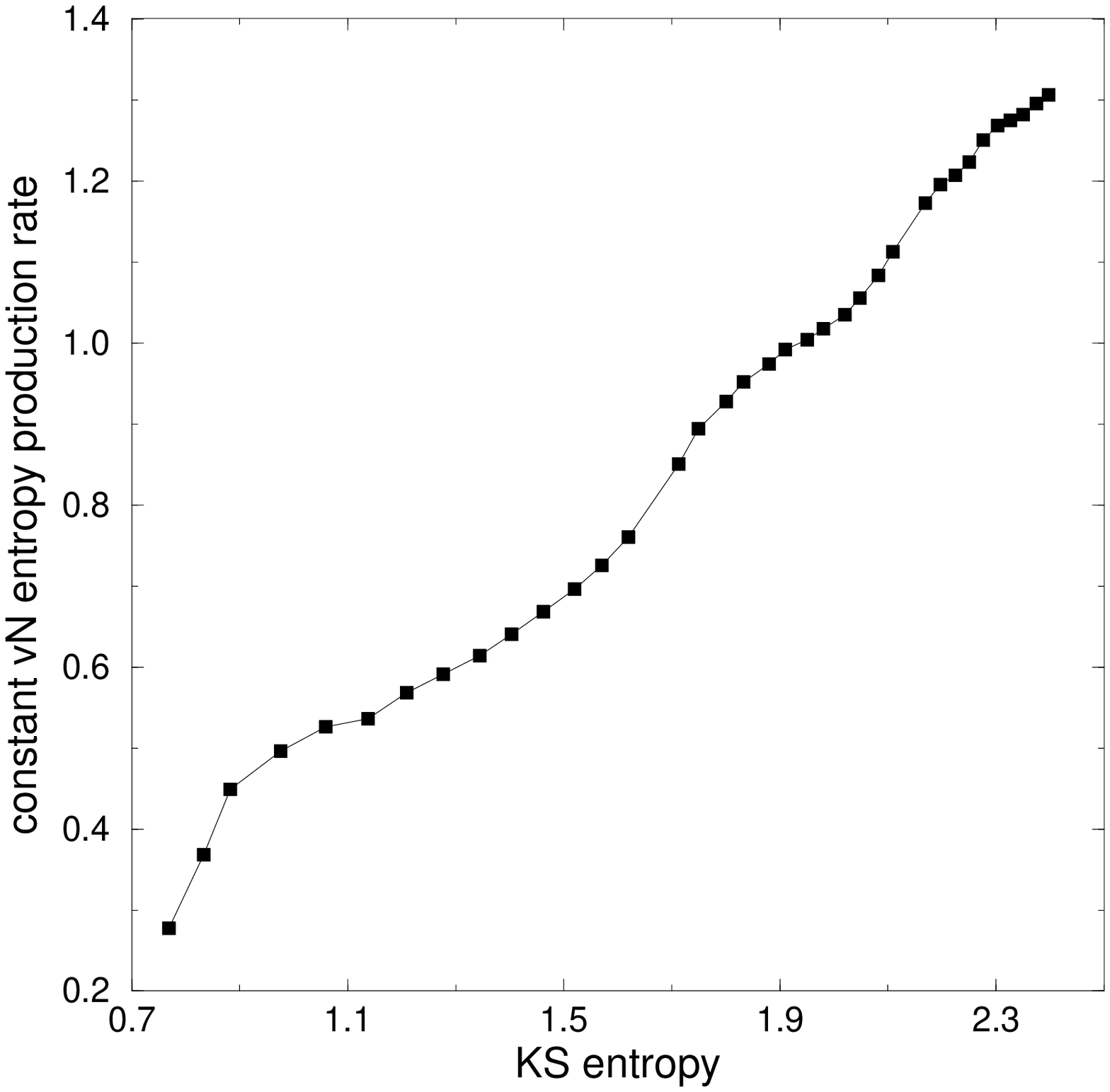}
\caption{The constant entropy production rate plotted against the 
corresponding local KS entropy. The starting point in phase 
space is $(0.5, h)$, where $h = 0.1 \gamma$. Here $\beta = 0.001$.}
\label{fig17}
\end{figure}

\begin{figure}[h]
\epsfxsize=6in
\epsfysize=6in
\epsffile{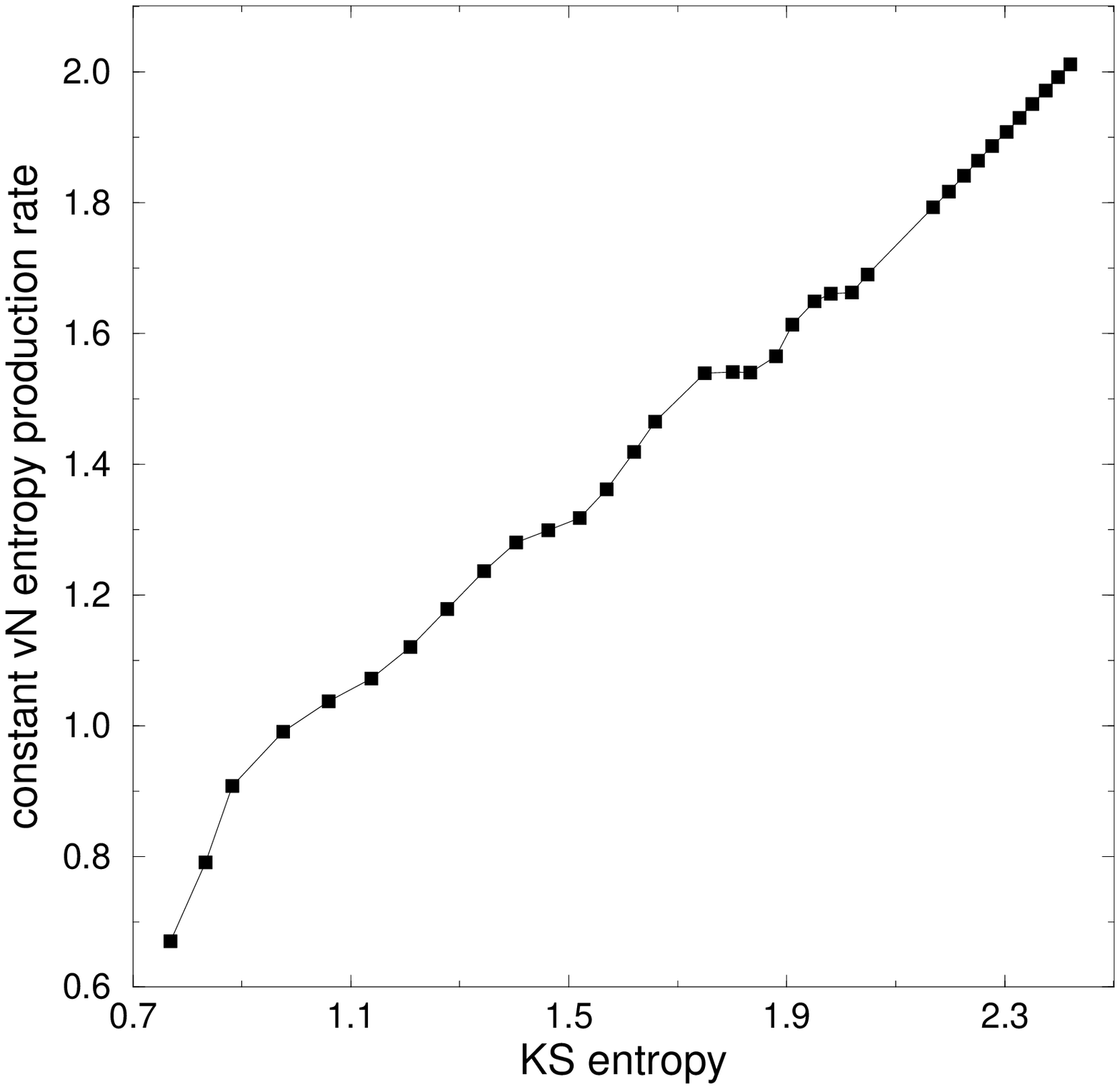}
\caption{The constant entropy production rate plotted against the 
corresponding local KS entropy. The starting point in phase 
space is $(0.5, h)$, where $h = 0.1 \gamma$. Here $\beta = 0.0001$.}
\label{fig18}
\end{figure}

In figures [\ref{fig16}], [\ref{fig17}] and [\ref{fig18}] we have done just 
that, but for values of the inverse temperature of $\beta = 0.1, 0.001$ and 
$0.0001$, respectively. The starting point was chosen as $(0.5, h)$. Thus, 
the same parameter values are used here for values of $K$ from 
$3.5$ to $ \approx 25$ as were used in figures [\ref{fig9}], [\ref{fig10}] and 
[\ref{fig11}] for $K = 3$. Figure [\ref{fig9}], especially, should give 
one an indication of the increased effectiveness of a larger environmental 
temperature in destroying the dynamical localization. This is manifest in 
these figures. As the temperature is increased we see a transition from 
quite a poor linear relationship in figure [\ref{fig16}] with large
fluctuations due to the existence of a localization effect (especially for 
smaller $K$ values) to obviously much better linear relationships in figure 
[\ref{fig17}] and [\ref{fig18}]. The fluctuations, though still present, 
are much reduced at these temperatures since there is now little or no 
localization to inhibit the spread of the evolving wave packet in 
the momentum direction. 
Once again we see that increasing the temperature has the effect of raising 
the slope.

\section{Conclusion}
\label{section5}

We have shown that the rate of von Neumann entropy production for an initial 
state of the open quantum kicked rotor increases approximately linearly 
with the {\em local} KS entropy of a corresponding classical distribution. 
This approximation becomes more exact as the strength of the environment 
becomes great enough to destroy on a short enough timescale those quantum
coherences which lead to dynamical localization in the absence of an
environment. However, even in that ``transition" region when the localization
is just noticably destroyed, linear behaviour is
observed, albeit with localization-induced fluctuations superimposed.      

We have also shown using this model, but believe it to be generally true given 
its prototypical nature, that: {\em a)} it is generally the case that the 
more classically chaotic a system is, the greater the rate its quantum 
counterpart 
will lose information to a perturbing environment; {\em b)} all system and 
environmental parameters being equal, different entropy production rates 
can still be seen depending on where one places the initial state in phase
space. These locations correlate excellently with the classical mixed phase
space structure, and finally; {\em c)} the constant rate of entropy increase 
depends very much on the nature and strength of the perturbing environment.

We note that observation {\em b)} here is not unique to the approach of
coupling quantum chaotic systems to a thermal environment, as has been done
here, or of otherwise perturbing it \cite{Zarum1,Zarum2}. Indeed, recently 
Mirbach and Korsch \cite{Korsch} have introduced a phase space entropy-like 
quantity, $S(p,q)$, which measures the (time averaged) spreading of mimimum 
uncertainty wave packets centred initially at $(p,q)$. No environment is 
introduced in their approach. Contour plots of this quantity show 
increasing similarities to both the classical phase space structure of 
Poincar\'{e} sections of a harmonically driven rotor, and to a suitably 
defined {\em classical} phase space entropy, as $\hbar$ is decreased. 
However, no connection is made with standard quantities used in the study 
of chaos, such as the KS entropy considered here.     

These results, we speculate, provide good criteria by which to distinguish 
between the quantum mechanical counterparts of regular and chaotic classical 
motion in general, even in multi-dimensional systems.

\ack

Paul A. Miller would like to thank the King's College London Association 
(KCLA) for a postgraduate studentship. We also thank Raphael Zarum for 
useful discussions.

\newpage


\end{document}